\documentclass[floatfix,aps,prb,twocolumn,groupedaddress,showpacs]{revtex4}

\usepackage{graphicx}

\begin{document}

\bibliographystyle{apsrev}

\title{Stripe as an effective one-dimensional band of composite excitations}
\author{A. L. Chernyshev}
\affiliation{Oak Ridge National Laboratory, P.O. Box 2008, Oak Ridge,
Tennessee 37831} 
\altaffiliation[Also at ]{Institute of Semiconductor
Physics, Novosibirsk, Russia.}

\author{S. R. White}
\affiliation{Department of Physics and Astronomy, University of California, 
Irvine, California 92697}

\author{A. H. Castro Neto}
\affiliation{Department of Physics, Boston University, Boston,
Massachusetts 02215}

\date{\today}

\begin{abstract}
The microscopic structure of a
charge stripe in an antiferromagnetic
insulator is studied within the $t$-$J_z$ model using analytical and
numerical approaches. 
We demonstrate that a stripe in an antiferromagnet
should be viewed as a system of composite holon-spin-polaron excitations 
condensed at the self-induced antiphase domain wall (ADW) of the
antiferromagnetic spins.  
The properties of such excitations are studied in detail with
numerical and 
analytical results for various quantities being in very
close agreement. 
A picture of the stripe as an effective one-dimensional
(1D) band of such excitations is also in very good agreement 
with numerical data. These results emphasize the primary role
of kinetic energy in favoring the stripe as a ground state. A 
comparative analysis suggests the effect of pairing and collective
meandering on the energetics of the stripe formation to be secondary. 
The implications of this microscopic picture of fermions
bound to the 1D antiferromagnetic ADW for the 
effective theories of the stripe phase in the cuprates are discussed.
\end{abstract}

\maketitle
\section{\label{intro}Introduction}

Strongly-correlated  models of the CuO$_2$ planes of high-T$_c$
superconductors continue to attract much attention due to the belief that
most of the physics in the cuprates is governed by strongly
interacting, purely electronic degrees of freedom \cite{pines}.
Microscopic studies of Hubbard and $t$-$J$ models have
been successful in explaining the $d$-wave character of the
pairing mechanism
\cite{Scalapino1,Poilblanc1,Kagan,BCDS} 
and other experimental
results, such as narrow low-energy bands
in the angle-resolved photoemission 
for the undoped systems. \cite{Nazarenko,me,narrow} More recent
interest in 
these models has been boosted by the discovery of stripes, or
spin and charge inhomogeneities, in high-T$_c$
materials.\cite{stripesExp,stripesExp1}
Generally, the strongly correlated $t$-$J$ and Hubbard models in
the non-perturbative regime ($J\ll t$ or $U\gg t$) are 
difficult to approach analytically, although
some advances have been achieved in solving them. 
Because of this,
numerical methods have been responsible for much of the
progress in the understanding of these
models.\cite{Dag,Prokofiev,Steve_Ian,Pakwo} 
Moreover, such numerical
studies have become a very important test of the ability of
theoretical approaches to describe the stripe and other
low-energy phases\cite{Lee} in cuprates.

While the striped phase was
anticipated from mean-field solutions of the Hubbard
model\cite{mf1}, probably the most convincing evidence of 
stripe-like ground states has been
provided by Density Matrix Renormalization Group (DMRG) studies of
the $t$-$J$ model in large clusters in the range
of parameters relevant to real systems.\cite{stripedmrg}
However, some other numerical approaches raise 
the question that the stripes seen in DMRG might be the
result of finite-size effects \cite{HM}. 
Another aspect of the problem is that
numerical studies alone do not directly answer
questions on the origin of stripes.
Ideally, one would wish for a theory which would closely agree
with the numerical data on all essential aspects, thus providing a
definite physical answer on how the stripes are created and what are
the excitations around this state. The search for such a case
has motivated our present work and that is the way we unify our
approaches here.\cite{SCES}

In this work we attempt to integrate some of the earlier ideas on the
$t$-$J$ model physics with the newer trends and phenomenology which
have appeared due to stripes. We approach the problem 
using a comparative study of the stripe in an antiferromagnetic 
insulator by DMRG and an analytical technique,
within the framework of the $t$-$J_z$ model.
Our numerical study utilizes DMRG in large $L_x\times L_y$ clusters of up
to $11\times 8$ sites, using various boundary conditions. 
The analytical method is a
self-consistent Green's function technique developed
earlier\cite{CP,CCNB}, which accounts for
the retraceable-path motion of the holes away from stripe. 
We demonstrate that the stripe in an antiferromagnet (AF)
should be viewed as a system of composite holon-spin-polaron excitations 
condensed at the self-induced antiphase domain wall ADW.

The $t$-$J$ model has long been seen as
a natural model for the description of the charges and spins in a
doped AF \cite{Anderson}. 
The single- and two-hole problems within the model have been studied
extensively using  different
analytical schemes and 
numerical approaches in small clusters
\cite{Dag,Prokofiev,Steve_Ian,Pakwo,KLR,SS,Eder1,YuLu,Reiter,Oleg,Barnes,BCS1,Steve_Doug,Hamer}. Some attempts to generalize the conclusions of
these studies on the nature of the many-hole ground-state have also been
made \cite{BCDS,Schrieffer,SS1,CLG}.  
Very good agreement between the numerical and analytical  studies 
for these problems has been achieved within the spin-polaron paradigm
\cite{Dag,CP,BNK,CLG}.
This essentially quasiparticle  picture describes the single-hole
excitation as a hole strongly dressed by the ``string''-like spin
excitations. It was found 
that two such quasiparticles tend to form a bound-state of 
$d$-wave symmetry.\cite{Dag,SS,Barnes,BCS1,Steve_Doug,Hamer,Ed1,Sushkov1}
The generic reason for the absence of the $s$-wave pairing is the
magnon-mediated exchange which generates a repulsion in the $s$-wave channel.
\cite{Scalapino1,Kagan,DJS} In the
$t$-$J$ model, vertex corrections are suppressed 
\cite{KLR,Sushkov2} and such a repulsion
is strong. Then the attraction in 
higher-order harmonics leads to the bound states  of higher
symmetry\cite{Sushkov3}. Attempts to integrate out the spin background
and to 
reformulate the $t$-$J$ model as an effective model 
for quasiparticles with a narrow band ($\sim 2J$) and an
interaction repulsive in the $s$-wave channel and 
attractive in the $d$-wave channel have been made,
\cite{BCDS,Sushkov1,SS1,CLG} assuming the antiferromagnetic  
correlation length  to be the largest scale in the problem. 
Phase separation in such a model in a physical range $t\gg J$ seems to be
unlikely since the pair-pair interaction  should also be repulsive, 
which agrees with the numerical data \cite{Poilblanc3}. 
The ground state in this model would be a dilute ``gas'' of $d$-wave
spin-polaron pairs.  

Such a  generalization of the spin-polaron
picture to a finite concentration of holes relies on the assumption
that the antiferromagnetic background remains unchanged. 
However, it is well known that the ``feedback'' effect of holes on
the antiferromagnetic background is important. Aside from
Hartree-Fock treatments of the Hubbard model \cite{mf1} which showed
stripe-like domain wall solutions, other studies of the $t$-$J$
model in the low-doping regime have indicated instabilities of the
antiferromagnetic order.\cite{instab} 
These instabilities were thought to lead
towards spiral\cite{spiral}, stripe-like spiral\cite{Dombre}, or
spin-liquid\cite{spin_liquid} states. Earlier numerical works in the
small $t$-$J$ clusters, Ref.~[\onlinecite{Prelovsek_old}], have
demonstrated stripes in the ground state which were also domain walls
in the N\'eel AF. With the mounting evidence from
experiments\cite{stripesExp,stripesExp1} and from DMRG numerical
data\cite{stripedmrg} the idea of {\it topological} doping\cite{KE}
has flourished. The spontaneously created ADW's have been widely
considered as the topological alternatives to the homogeneous N\'eel
background.\cite{Emery_big,AHCN,Antonio_Hone,stripes}

Thus, the many-hole ground state has turned out to be very different
from that for a few holes. In order to understand the nature of the
charge excitations 
in this phase one needs to reconsider the
single-particle problem around this ground state with different
topology.\cite{Wilczek} The one-dimensional (1D) 
character of the charge stripes
has led to a number of attempts to generalize the physics of strictly
1D systems, where the excitations are holons and spinons, to higher
dimensions.\cite{Zaan1,1D_to_2D} 
On the other hand, there is a growing understanding that the stripes
are the outcome of the same tendencies which are seen already for the
single-hole problem\cite{Martins}, and that the charge excitations in
the stripe phase may still have lots in common with the spin
polarons.\cite{CCNB,Wrobel_stripe} 

Note that topological doping generally refers to the introduction 
of dopants into the topological defects of a field theory, which  
are the field configurations that interpolate between
different vacua of the problem. For problems with high
symmetry, such as the  $SU(2)$ Heisenberg model, topological defects
are continuous distortions of the order parameter.\cite{S_abanov}
The ADW's are the topological defects for 
systems with lower $Z_2$ symmetry, such as the Ising or the 
anisotropic $t$-$J_z$ model 
(and also models for polyacetylene\cite{SSH}).  While the magnetism in
cuprates is very well described by the $SU(2)$ symmetric 
models\cite{Kastner} the 
experimental finding of stripes indicate that topological doping 
corresponds to topological defects of lower symmetry. 
Although the reason for this lowering of the symmetry is, most probably,
dynamical in nature and is still not clearly understood, it gives us
confidence that the $t$-$J_z$ model is the right starting point
for the description of these systems.

As long as one is concerned with
the short-range physics of the charge and spin excitations,
the isotropic $SU(2)$ $t$-$J$ and
anisotropic $t$-$J_z$ models lead to similar results, as is well 
known from earlier studies.\cite{Dag} 
Roughly, the hole motion is fast and the spin relaxation is slow when
$t\gg J$. Therefore, in the fast timescale the hole moves in the
background of essentially static, staggered spins.
We will show below that the stripe
phases obtained numerically for the $t$-$J_z$ model are virtually
identical to those in the $t$-$J$ model studied
before.\cite{stripedmrg} 
It has also been concluded, based on the Ginzburg-Landau functional
approach, that the antiphase shift of the antiferromagnetic order
parameter must originate from some short-range physics.\cite{Pryadko} 
The rigidity
of the $\pi$-shift of the antiferromagnetic phase across the domain
wall in both numerical and experimental studies also argues for the
short-range genesis of the stripes. 

In this paper we, therefore, study analytically and numerically 
the system of holes at the ADW in the anisotropic $t$-$J_z$ model. 
While we restrict
ourselves to the study of the strongly anisotropic limit of the basic
$t$-$J$  model, we believe that the results of our study are much more
generic since the strongly correlated nature of the problem is preserved.

Technically, switching off the transverse spin fluctuation 
has numerical and analytical advantages. The numerical advantages are
twofold. First, the sizes of the system which can be handled by the DMRG
method are considerably bigger. Second, one can think of the boundary
conditions as a way of stabilizing  ground states of
different symmetries. In our case the choice would be between the
state with and the state without an ADW. Then, one can consequently dope 
holes in the stripe (ADW) 
configuration and study the evolution of the
properties of the system as a function of doping, starting from a very
dilute limit. It is worth noting the boundary conditions in our
case work as a very gentle  instrument of controlling the symmetry of
the ground state {\it without} affecting the wave-functions of the
states themselves. The analytical advantage is the treatability of the
problem. The analytical part of this work largely relies on a
previous study by two of us and Bishop, Ref.~[\onlinecite{CCNB}],
where we have calculated the Green's function
of the charge excitation at the ADW by a method 
which goes beyond the limits of mean-field or perturbation theory.

The general conclusion of the present study is that the stripe should
be considered as a collective bound state of the holes with an
ADW. In such a system the excitations are composite 
holon-spin-polarons which populate an effective 1D band. 
This picture is in very good agreement with the numerical results
and provides insight into the problem of the stripe phase in
cuprates. 

This paper is organized as follows. Section \ref{Analytics} describes
various aspects of our analytical approach in detail. Section \ref{Numerics}
describes the DMRG method. Section \ref{results} presents the results and
comparison. 
Section \ref{conclusions} lists our conclusions. 

\section{Analytical approach}
\label{Analytics}

Our starting point is the $t$-$J_z$ model which is given by:
\begin{eqnarray}
\label{H_tJ}
{\cal H} = -t\sum_{\langle ij\rangle\sigma}(\tilde{c}^\dagger_{i\sigma}
\tilde{c}_{j\sigma}+{\rm H.c.})+ 
J\sum_{\langle ij \rangle} \bigl[S_i^z   S_j^z-
\frac{1}{4}N_iN_j \bigr]\ , 
\end{eqnarray}
where $t$ is the kinetic energy, $J$ is the antiferromagnetic
exchange, and $N_i=n_{i\uparrow}+n_{i\downarrow}$. 
All operators are defined in the space without
double-occupancy of the sites. 

\begin{figure}[t]
\includegraphics[width=3in,clip=true]{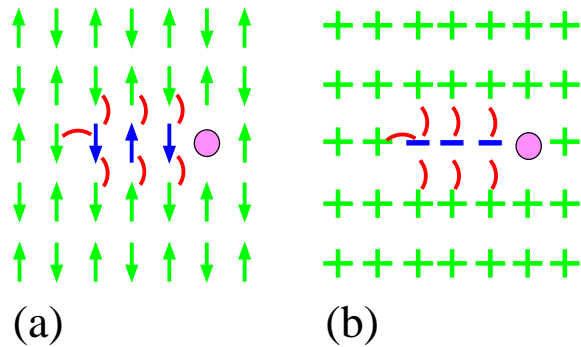}
\caption{\label{fig_0}(a) A hole followed by the ``string'' of spin defects
in a homogeneous AF. (b) same as (a), ``$+$'' and ``$-$'' denote
the sign of the staggered magnetization $M_i=(-1)^i S^z_i$.
Arcs denote ``wrong'' (ferromagnetic) bonds.}
\end{figure}

The single-hole problem for the $t$-$J$ ($t$-$J_z$) model in a
homogeneous antiferromagnetic background is well studied
with analytical results and numerical data being in very good 
agreement \cite{Dag,CP,CLG}. 
The charge quasiparticle is understood as a spin
polaron, i.e. a hole dressed by strings of spin excitations
\cite{BNK}. It is also often expressed as that
the hole movement in a homogeneous antiferromagnetic background is
frustrated because of the tail of misaligned spins following the
hole, see Fig. \ref{fig_0}. The idea that an ADW can be a more favorable
configuration for holes relies on the fact that such a frustration of the
hole's kinetic energy can be avoided for a movement inside the wall,
such that the hole is essentially free in the 1D structure.
However, as we show below, the spin-polaron aspect of the physics of
the charge carrier remains very important in the stripe phase as
well. We will first consider the specifics of the hole behavior in the
inhomogeneous antiferromagnetic state (state with an ADW) and will
address  the 
spin-polaron aspect of the problem later.
For the detailed description of the spin-polaron formalism used in
this work we refer to Ref.~[\onlinecite{CP}].

\subsection{Holon in the domain wall}
\begin{figure*}
\includegraphics[width=1.6in]{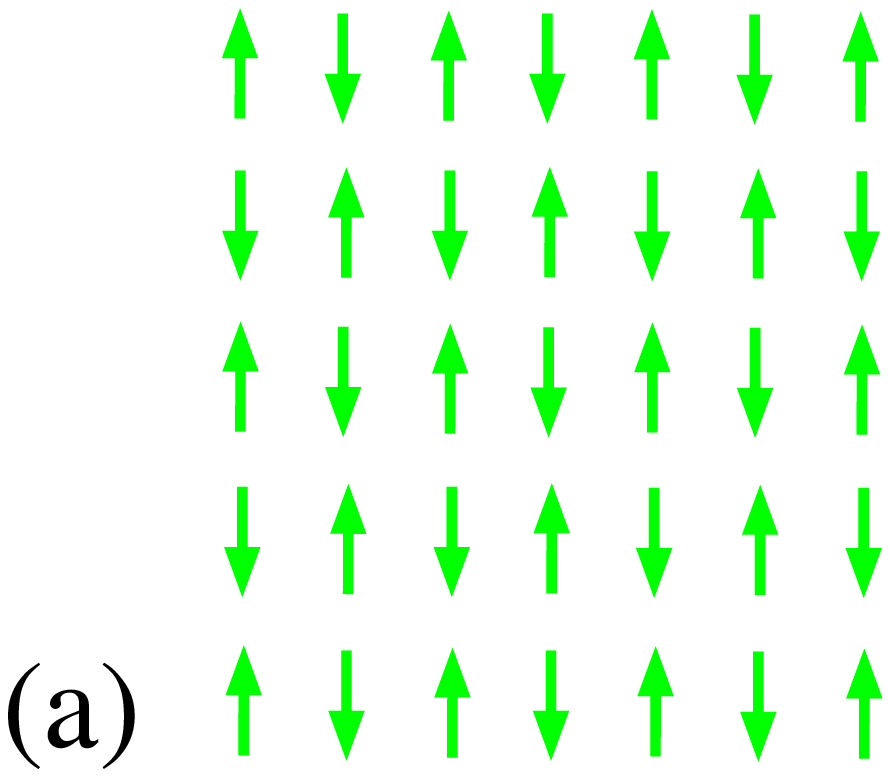} \hskip 1.5cm
\includegraphics[width=1.6in]{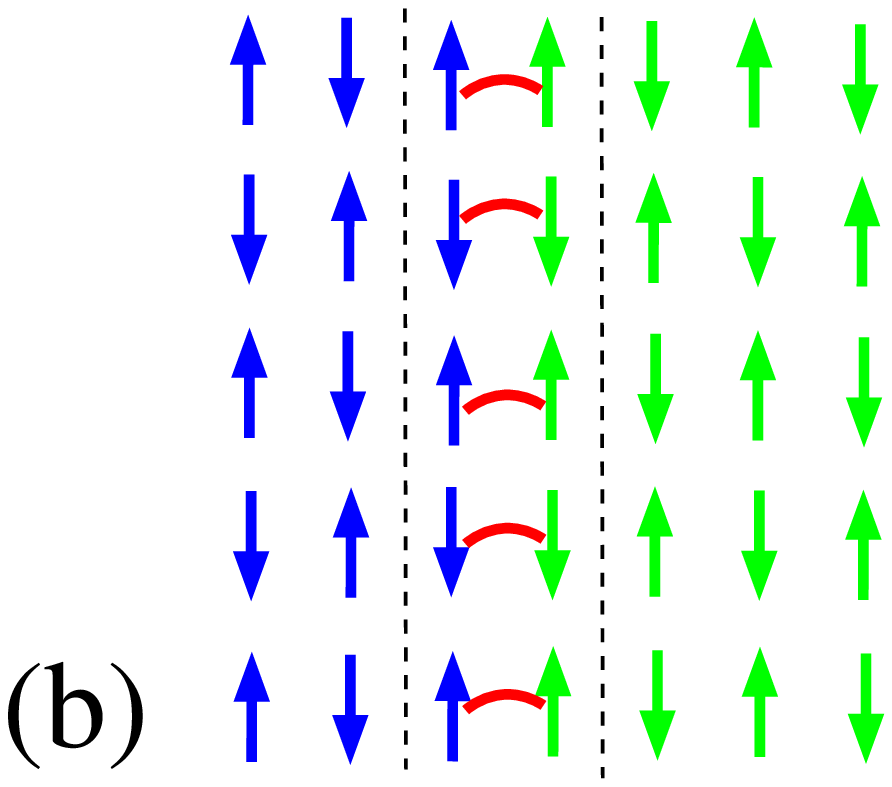} \hskip 1.5cm
\includegraphics[width=1.6in]{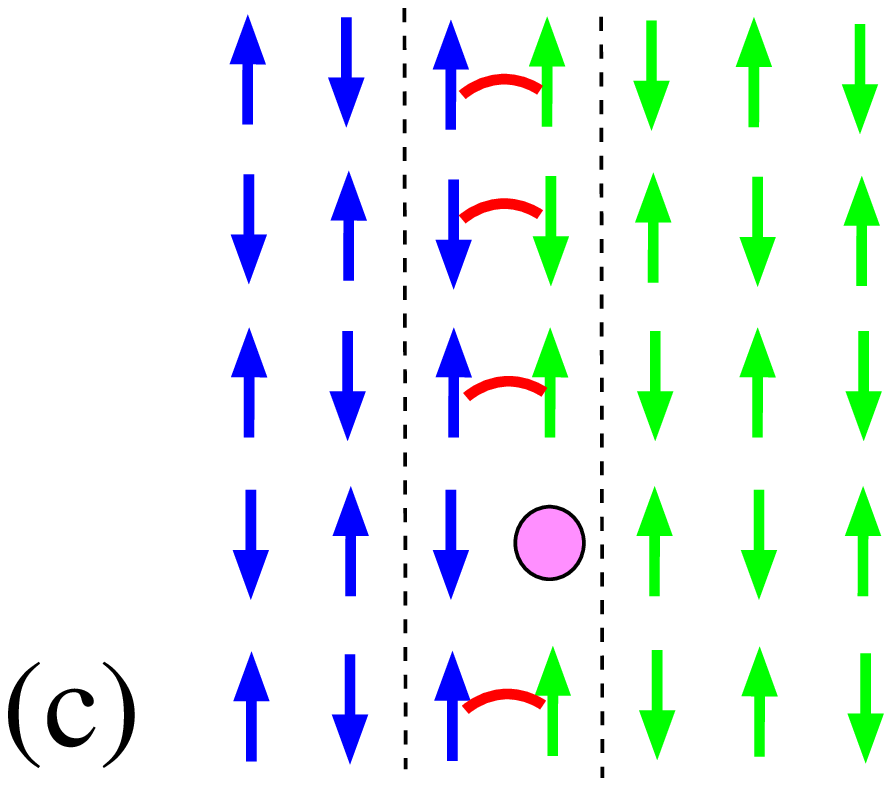}
\caption{(a) Homogeneous  N\'eel state, (b) two domains of an AF
with opposite staggered magnetization separated by the bond-centered 
ADW, (c) a static hole ($t=0$) is attracted to the ADW. 
Arcs denote ``wrong'' (ferromagnetic) bonds.}
\label{fig_1}
\end{figure*}

Since the stripe corresponds to  an
ADW in the spin background, one has to study the nature of charge
excitations at such a domain wall. Let us consider the empty system
first. The ground state is, of course, given by the simple N\'eel
configuration of spins, Fig. \ref{fig_1}(a). 
However, when the antiphase shift of the staggered magnetization is
created (enforced by the boundary conditions, for instance) the straight 
bond-centered domain wall is the lowest
energy state, Fig. \ref{fig_1}(b). It has the energy $E^J_{bond}=J/2$
per unit length, which is lower, for example, than the corresponding
energy for the site-centered domain wall, $E^J_{site}=3J/4$. We remark
here that the $SU(2)$ Heisenberg spins would prefer a continuous untwist
from one end of the crystal to the other without a sharp domain wall.
 
Now let us consider a single hole doped to the system. When the kinetic
energy is neglected ($t=0$) the lowest energy state is defined from 
simple bond counting. Evidently, the hole is attracted to the domain wall 
since the potential energy ($J$-term) is
lowered when the hole removes the ``wrong'' bond, Fig. \ref{fig_1}c. 
\begin{figure}[b]
\includegraphics[height=2in,clip=true]{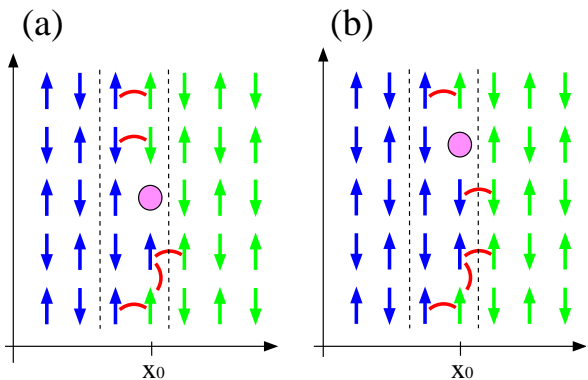}
\caption{``Longitudinal'' hole motion along the ADW. The $x=x_0$ line
can be considered as an Ising chain. Arcs denote ``wrong''
(ferromagnetic) bonds.}
\label{fig_2}
\end{figure}

\begin{figure}[b]
\includegraphics[width=3in,clip=true]{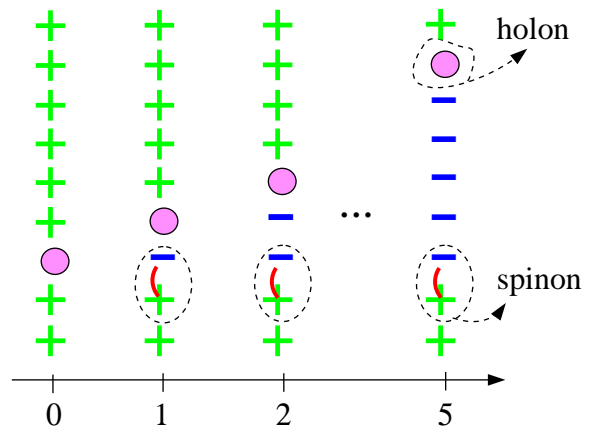}
\caption{Propagation of the hole in the Ising chain. Numbers 
indicate the amount of hoppings made by the hole away from its 
origin, ``$+$'' and ``$-$'' represent the sign of the staggered
magnetization.}
\label{fig_3}
\end{figure}

When the kinetic energy is taken into consideration the following
observation 
can be made. If one restricts the hole motion to one side of the ADW
``ladder'' of defects (along the $y$-axes, $x$ is fixed at $x_0$ in Fig.
\ref{fig_2}) 
the problem
is identical to the hole motion in the 1D Ising chain.\cite{1D} 
That is, one can see that after the
first step a 1D spin defect (spinon) is created and then the motion of the 
hole does not cause any further disturbance in the spin background, 
Figs. \ref{fig_2}(a,b). The hole simply rearranges the ``wrong'' bonds while
moving.
Since the charge does not carry any ``memory'' about the
spin of the place where it was created this excitation is a holon. 
Notably, when the spinon and holon are separated, they both
carry a ``kink''  or ``anti-kink'' of the staggered magnetic order;
in other words, they are zero-dimensional ADWs in the 1D chain problem,
Fig. \ref{fig_3}.

It is more visually convenient to use staggered
magnetization $M_i=(-1)^i S^z_i$ instead of the on-site magnetization
to emphasize the  opposite direction of the order parameter in the antiferromagnetic
domains of spins. We, therefore, will often use ``$+$'' or ``$-$'' 
instead of
the actual direction of the spin.

\begin{figure*}
\includegraphics[width=1.6in]{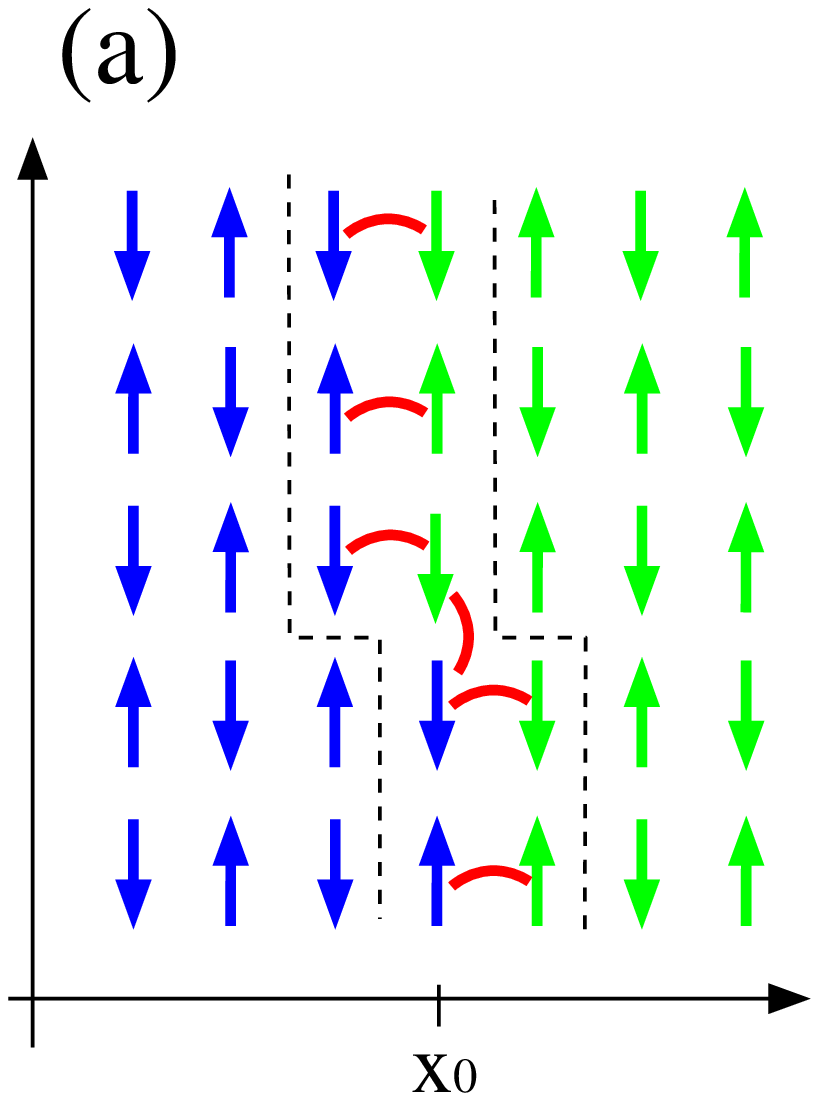} \hskip 1.5cm
\includegraphics[width=1.6in]{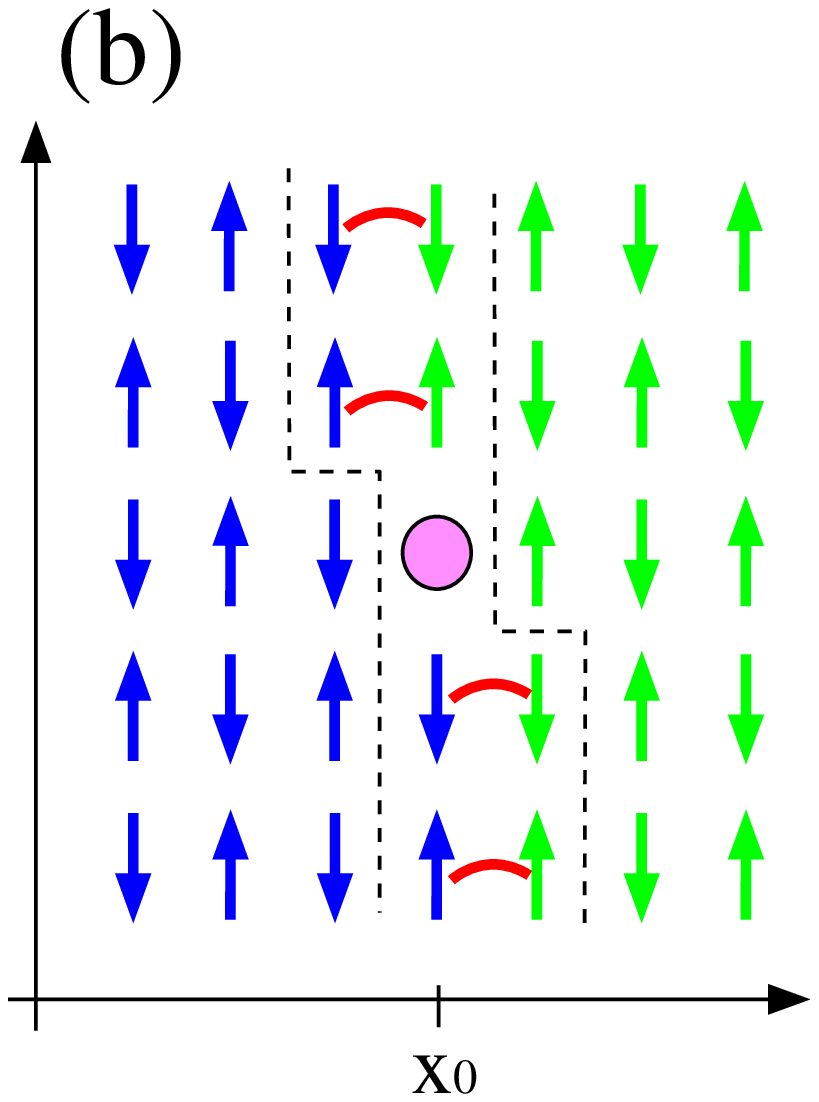} \hskip 1.5cm
\includegraphics[width=1.6in]{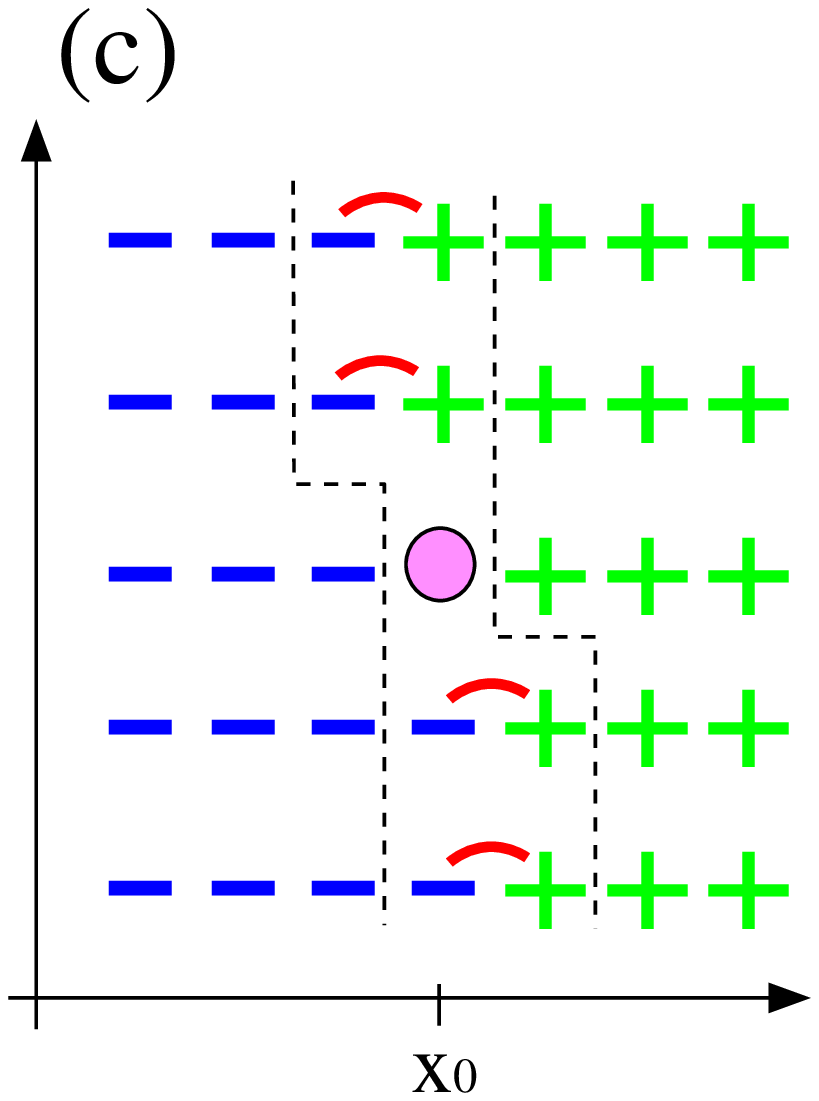}
\caption{(a) An empty ADW with the ``wiggle''. For this 
configuration $x=x_0$ line is an Ising chain with the single spinon.
(b) A hole created at the place of the spinon is a holon. (c) same 
as (b) with ``$+$'' and ``$-$'' showing the sign of the staggered
magnetization. Arcs denote ``wrong'' (ferromagnetic) bonds.}
\label{fig_5}
\end{figure*}

Since the spinon in our case is a finite-energy excitation  
($E_{spinon}=J/2$), in the $t\ll J$ limit the hole will be always 
bound to its site of origin. One can think of several possibilities 
to avoid the spinon creation in order 
to focus on the properties of the charge
excitation only. One can ({\it i}) assume that the crystal is 
finite (semi-finite) and create a hole at one of the ends of the chain, 
then the propagation of the holon along the chain is free, 
({\it ii}) create a pair of holes 
at the nearest neighbor sites and then consider their
motion independently, in this case both 
the kink and anti-kink are carried by holons, and ({\it iii})
start with the empty domain wall with a ``wiggle'', 
one half of it misplaced 
by one lattice spacing from the other half along the $x$-axis. This way
one has an
extra
``wrong'' bond in it, Fig. \ref{fig_5}(a), which is equivalent to 
having a 1D chain along the $x=x_0$ line with the single spinon. Then the 
hole creation at one of the sites forming the spinon is identical to 
the creation of a free holon, Figs. \ref{fig_5}(b,c). 
The purpose of these manipulations is to show that the single 
hole motion along the ADW can be made free provided that the spin 
environment ensures the holonic nature of the charge excitation.
These considerations are by no means new and were discussed in
Refs.~[\onlinecite{Oleg_T,Erica}]. We suggest
calling the free motion of the hole within the stripe
a ``prepared-path'' motion in accord with 
the ``retraceable-path'' motion for the spin polaron.

As it follows from the above consideration, the wave-function 
of the single holon cannot be simply written as a result of 
an action of a single annihilation operator on some unique 
ground-state wave-function, since it requires a rearrangement of 
the (semi-) infinite amount of spins. However, if the ground-state 
is ``prepared'', as in the case ({\it iii}), one can still keep the 
formal similarity with the conventional single-particle creation.
In other words, if the holon is to be created at the site $i$, its
wave-function can be written as:
\begin{eqnarray}
\label{psi}
&&|1_i\rangle=c_{i,\sigma_i}|\psi_{0,i}\rangle = (-1)^{i-1}
\prod_{j<i} c^\dag_{j,\sigma_j}\prod_{j>i} c^\dag_{j,{\bar\sigma_j}}
|0\rangle\nonumber\\
&&\phantom{|1_i\rangle}\equiv(-1)^{i-1}|\dots \uparrow\downarrow 
\uparrow\downarrow \mbox{\footnotesize  $\bigcirc$}
\uparrow\downarrow \uparrow\downarrow \dots\rangle\\
&&\phantom{|1_i\rangle}\equiv(-1)^{i-1}|\dots ++++\,
\mbox{\footnotesize  $\bigcirc$}
\, ---- \dots\rangle \ , \nonumber 
\end{eqnarray}
where the Hilbert space at each site is restricted to single 
occupancy, ${\bar\sigma}=-\sigma$, 
$\sigma_j=\uparrow (\downarrow)$ if $j\in A(B)$ 
sublattice, $|0\rangle$ is the vacuum state, ``$+$'' and ``$-$'' 
denote
the domains with staggered magnetization ``up'' and ``down'',
respectively. This definition suffices for our consideration.
 
The above mentioned ``restriction'' of the hole motion along the 
ADW can be formally expressed as a separation of the kinetic energy 
term in the Hamiltonian (\ref{H_tJ}) in two pieces:
\begin{eqnarray}
\label{H_t}
{\cal H}_t={\cal H}^{\parallel}_t+{\cal H}_t^\perp\ ,
\end{eqnarray}
where the first term (``longitudinal'') 
includes only 1D motion along $y$-axis for 
$x=x_0$, and the second term (``transverse'') includes the rest.
Evidently,
\begin{eqnarray}
\label{H_t_psi}
{\cal H}^{\parallel}_t|1_i\rangle
= \, t\left(|1_{i-1}\rangle+|1_{i+1}\rangle \right) \ ,
\end{eqnarray}
and thus 
\begin{eqnarray}
\label{H_t_psi_k}
{\cal H}^{\parallel}_t|k\rangle=2t\, \cos k \, |k\rangle \ ,
\end{eqnarray}
where 
\begin{eqnarray}
\label{psi_k}
|k\rangle=\sum_i e^{ikr_i}|1_i\rangle \ .
\end{eqnarray}
Consequently, the single-particle ``bare'' Green's function 
of the free spinless fermion (holon) propagating along the 
ADW with simple tight-binding dispersion can be written as 
\begin{eqnarray}
\label{G_0}
G_{x_0}^{(0)}(k_y,\omega)=
\langle k_y |\frac{1}{\omega-{\cal H}_t^{\parallel}}
|k_y\rangle=\frac{1}{\omega-2t\cos k_y +i0}\ ,
\end{eqnarray}
where the zero of energy is set at the energy of the lowest static 
($t=0$) single-hole state $E_0=\langle {\cal H}_J\rangle_{t=0}
=E_{Ising}+(L_y-1)J/2+2J$, and $L_y$ is the size of the plane in the
direction of the stripe.  
The holon band minimum is located at $k_y=\pi$, and the 
index $x_0$ corresponds to the $x$-coordinate of the stripe.
\begin{figure}[b]
\includegraphics[angle=270,width=3in,clip=true]{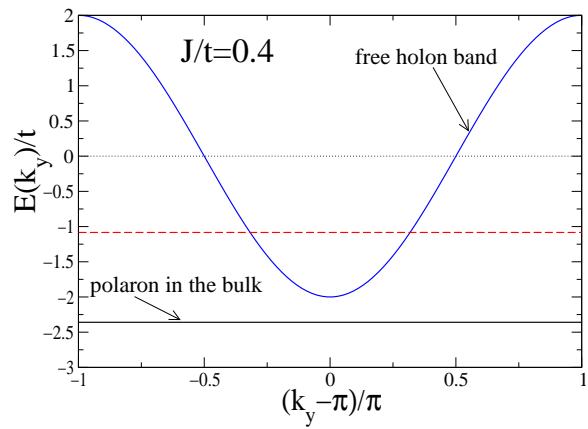}
\caption{The free holon band ($E_k=2t\cos k$, solid curve), energy 
of the spin polaron in the bulk (solid line), and energy per hole
in the 1D holon band at half-filling, $E_{1/2}=J/2-4/\pi$ (dashed line),
are shown.}
\label{fig_6}
\end{figure}
\begin{figure*}
\includegraphics[width=1.6in]{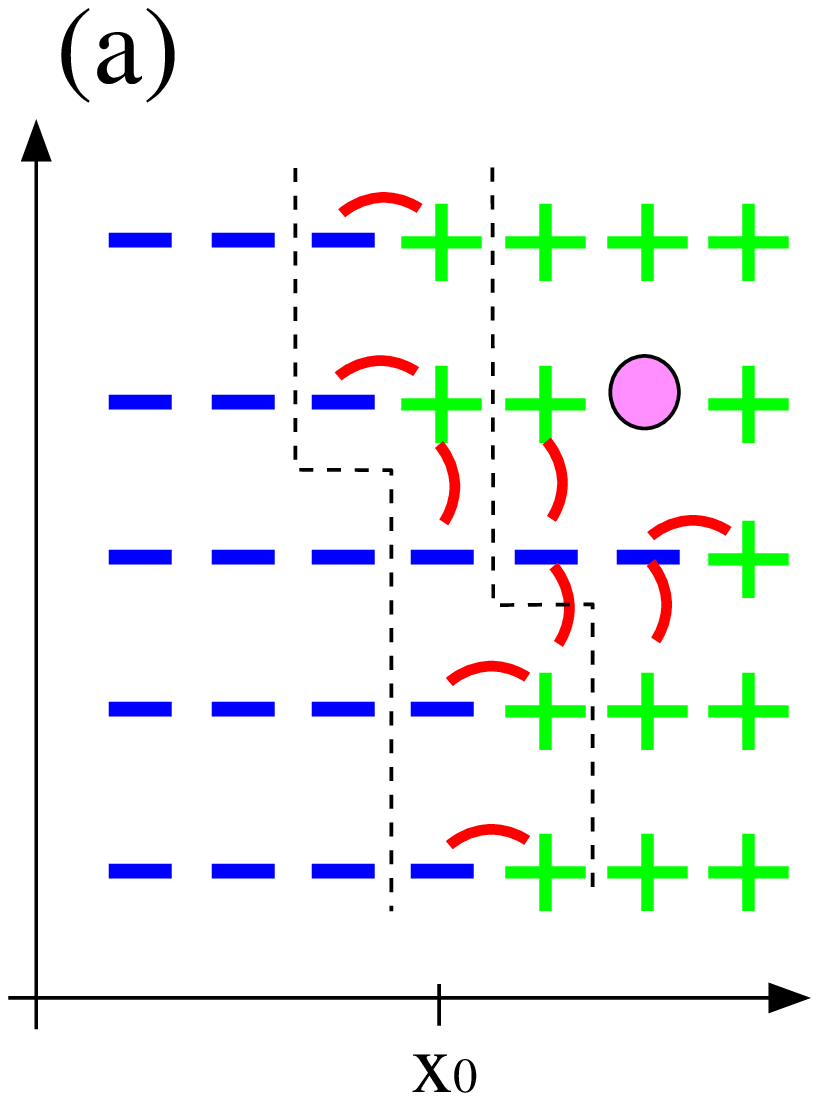} \hskip 1.5cm
\includegraphics[width=1.6in]{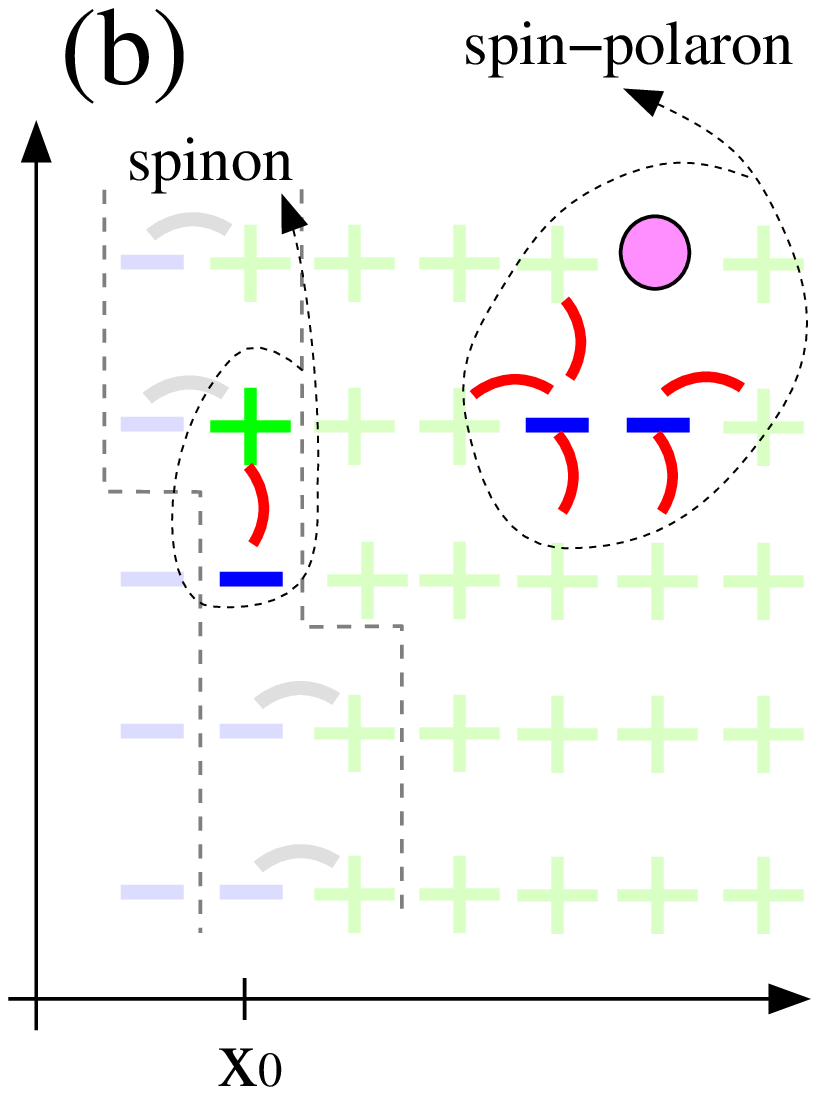} \hskip 1.5cm
\includegraphics[width=1.6in]{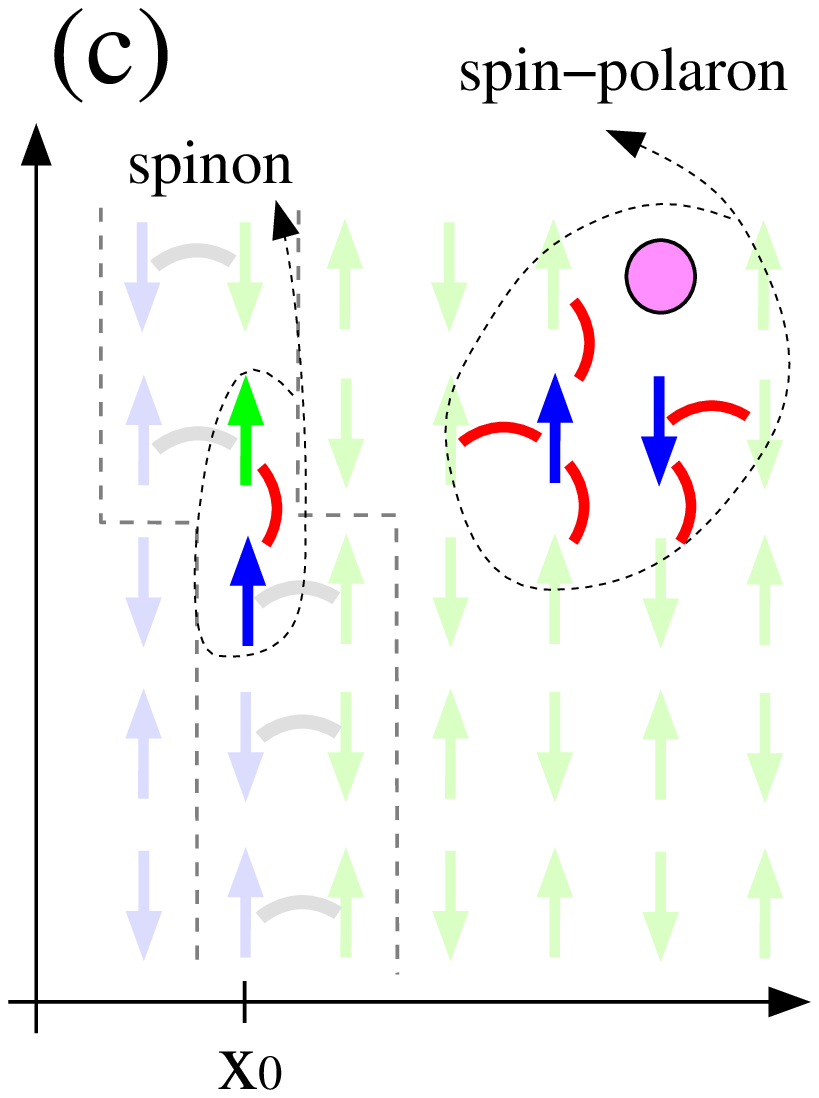}
\caption{(a) An example of a ``string''
generated by the ``transverse'' hole movement.
Notably, the first defect in the ``string'' is a spinon in
the $x=x_0$ Ising chain. (b) A schematic result of the hole departure from
the ADW, ``$+$'' and ``$-$'' showing the sign of the staggered
magnetization as before. (c) same as (b) with the actual spin
directions shown.}
\label{fig_7}
\end{figure*}

This observation 
that the hole can avoid the frustration of the antiferromagnetic
background in the presence of an ADW in comparison with the 
homogeneous N\'eel state where the hole motion always leads to 
string-like spin defects has been known since the discovery
of the stripe phases. However, the sufficiency of this effect
{\it alone} to justify the stripe formation has been exaggerated. 
Roughly speaking,  this argument
is gathered from the  unphysical limit of the model,  $t\ll J$,
where the {\it kinetic} energy of the hole 
is indeed lower in the state with an ADW 
($E^{ADW}_{kin}\sim -t$) than in the homogeneous state with a spin
polaron ($E^{sp}_{kin}\sim -t^2/J$). However, the energy cost of the
domain wall, $E_{wall}\sim J$, per unit length overwhelms this gain
in kinetic energy in
this limit.  Moreover, the true ground state in the $t\ll J$ limit
is neither a stripe nor a homogeneous 
N\'eel state with holes, but rather the phase separated hole-rich and no-hole
states. 
In the physical limit $t\gg J$ the ``longitudinal''
kinetic energy of the holon at the bottom of the ADW 1D
band is $-2t$, while the energy of the hole (spin polaron) in the
``bulk'' (homogeneous N\'eel) is 
\begin{eqnarray}
\label{E_sp}
E_{sp}=-2\sqrt{3} t + {\cal O}(J^{2/3}t^{1/3}) \ .
\end{eqnarray}
Evidently, the ``unfrustrated'' kinetic 
energy alone is insufficient to compete with the energy of 
the homogeneous state. We further illustrate this statement below.

Figure \ref{fig_6} shows the energies of the spin polaron 
in the bulk and the dispersion
of the pure holon state at the ADW
for the realistic ratio of $J/t=0.4$.
In the $t$-$J_z$ model the dispersion of spin polarons is small
and we neglect it from this picture. One has to bear in mind that 
there is an (infinite) energy offset between these two lines: the energy 
of the ADW $E_J^{ADW}=L_y J/2$. This simply means that the single 
holon cannot compensate the price for the domain wall 
creation. Thus, the holon band must be ``filled'' to a certain level 
in order to reduce the energy. When we add more holes they 
 will fill the higher $k$-states in
the band. Assuming a rigid-band filling
the total energy per hole as a function of the 1D filling fraction
$n_{\parallel}$ can be calculated
\begin{equation}
\label{E_holon}
E_{tot}/N_h=\frac{J}{2}\left(\frac{1}{n_{\parallel}}-1\right)
-\frac{2t\sin\pi n_{\parallel}}{\pi n_{\parallel}} \ .
\end{equation}
which is infinite at $n_{\parallel}\rightarrow 0$, $E_{tot}=0$ 
for completely filled band $n_{\parallel}=1$, and has a shallow
minimum at some intermediate value of $n_{\parallel}$. For a chosen
value of $J/t=0.4$ this minimum is around $n^{min}_{\parallel}=0.32$. 
This lowest energy $E_{min}\simeq -1.255t$ is a bit lower than
the energy of the half-filled band $E_{1/2}=J/2-4t/\pi\simeq -1.07t$ 
shown in Fig. \ref{fig_6} by the dashed line. One can see that these
energies are more than $t$ higher than the energy of the 
spin-polaron system in the homogeneous N\'eel state 
$E_{sp}\simeq -2.37t$. Therefore, the energy
balance of a ``narrow'', 
strictly 1D stripe v.s. polarons is strongly against the stripe. 

\subsection{Transverse hole motion}

This illustration brings up the importance of the ``transverse''
part of the kinetic energy for the stripe formation. 
The ``transverse'' motion of a hole from the ADW, which 
includes all possible paths and not only those perpendicular to the
stripe, is by no means different from the ``string'' type of 
propagation in the homogeneous AF, compare Figs. \ref{fig_0} and 
\ref{fig_7}(a). That is to say that the charge excitation must
essentially regain its spin-polaron properties away from the domain
wall. Our Fig. \ref{fig_7}(a) shows an example of a ``string''
generated by such a transverse movement. It is well known that the 
hole can propagate by erasing the tail of ``wrong'' spins via the
so-called  
Trugman processes.\cite{Trug} There is an important qualitative feature of 
our case which makes it different from the 
homogeneous problem in this aspect. 
Since the excitation inside the ADW is a holon, 
that is $Q=1$, $S^z=0$ excitation, while the spin polaron is a 
``normal'' quasiparticle, $Q=1$, $S^z=\pm 1/2$, the conservation
of the quantum numbers requires the departure of the holon from the 
stripe to be always accompanied by the emission of the spinon 
($Q=0$, $S^z=\pm 1/2$). This is clearly the case as is shown in
Figs. \ref{fig_7}(b,c). 

\begin{figure}[t]
\includegraphics[width=2.5in,clip=true]{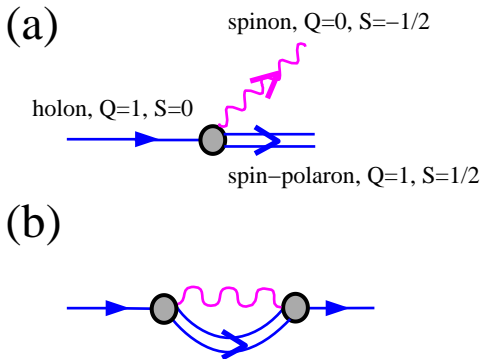}
\caption{(a) Decay of the holon into a spinon and a spin polaron. (b) 
The self-energy associated with such a decay.}
\label{fig_8}
\end{figure}

In other words, the ``transverse'' hole 
motion should be considered as a decay process of the 1D 
(ADW) charge excitation into a 1D spin excitation 
and a ``bulk'' charge excitation, Fig. \ref{fig_8}(a).
Since both the holon dispersion and the holon-spin-polaron coupling
are given by the same parameter $t$ such virtual decays will lead 
to a strong renormalization of the holonic energy band.

As shown in Eq. (\ref{G_0})
the ``bare'' Green's function of the hole residing at the ADW
 is the Green's function of a free spinless fermion
(holon) with tight-binding dispersion.
The renormalization of this Green's function is 
given by the self-energy schematically shown in Fig. \ref{fig_8}(b). 
It is well known in the single-hole problem for the $t$-$J$ model that
the self-consistent Born approximation, which is equivalent 
to the retraceable path approximation in our
case, accounts for the absolute
majority of such a renormalization.\cite{CP} 
This latter fact is related to the
effective analog of the Migdal theorem in this class of problems:
all first-order corrections to the hole-magnon vertex are zero because
of spin conservation.\cite{KLR} 
Corrections from the higher-order processes,
also known as Trugman paths,\cite{Trug} remaining beyond the
retraceable path approximation are negligible. 
In anticipation of the
further results we have to remark that in our problem the Trugman
processes are simply 
forbidden by the conservation of quantum numbers for the
holon decay. Therefore, the retraceable path approximation should work
even better due to the fact that the holon must emit the spinon
with the  energy $\sim J$. 
In other words, at least one spin
excitation is always created and thus the Trugman
paths belong to the same class as all other renormalization processes.

In our case the calculation of the self-energy in Fig. \ref{fig_8}(b) is
particularly simple since the spinon is a dispersionless excitation and
the double line corresponds to the spin-polaron Green's function known
from the previous studies \cite{CP}.
As in the case of the spin polaron, Ref.~[\onlinecite{CP}], 
the renormalization is coming from the retraceable path movements of
the hole away from ADW and back. 
The full Green's function is then given by
\begin{equation}
\label{G}
G_{x_0}(k_y,\omega)=\frac{1}{\omega-2t\cos k_y-\Sigma(\omega)+i0}\
, 
\end{equation}
where $\Sigma(\omega)$ takes the form of a continued fraction,
\begin{equation}
\label{S}
\Sigma(\omega)=\frac{(z-2)t^2}{\omega-\omega_1-\frac{(z-1)t^2}
{\omega-\omega_1-\omega_2-\dots}}\ ,
\end{equation}
$z=4$ is the coordination number,
$\omega_i$ is the energy of the $i$'th segment of the string, which
is equal to the number of ``wrong'' bonds ($J/2$ each) associated with
the segment, index $x_0$ corresponds to the $x$-coordinate of the stripe. 

The energy spectrum of the elementary excitations is given
by the poles of the Green's function Eq. (\ref{G}), therefore
one needs to calculate $\Sigma(\omega)$ and seek 
solutions of $E(k_y)-2t\cos k_y -\Sigma(E(k_y))=0$. 
The resulting effective 1D band for the composite holon-spin-polaron
excitation has been calculated in the previous work,
Ref.~[\onlinecite{CCNB}]. 
A standard simplification in the calculation of $\Sigma(\omega)$ 
is to assume that all $\omega_i$'s in Eq. (\ref{S}) are identical so
that the energy of the string is simply proportional to the length
of the path and is independent of the path of a hole.
This is a good approximation for the spin polaron
because only very few strings do not follow this rule. 
With this assumption the solution for the self-energy can be found in
a compact analytical form given by the ratio of the Bessel
functions \cite{Oleg}. If we take $\omega_1=J/2$
(energy of the spinon) and 
$\omega_{i>1}=J$ [two ``wrong'' bonds per segment of the string, see
Fig. \ref{fig_7}(a)], the self-energy is \cite{Oleg,CP}
\begin{equation}
\label{S1}
\Sigma(\omega)=\frac{2t^2}{\omega-J/2+\sqrt{3}t\Upsilon(\omega-J/2)}\ ,
\end{equation}
with $\Upsilon(\omega)={\cal J}_{-\omega/J}(r)/ 
{\cal J}_{-\omega/J-1}(r)$, ${\cal J}_\nu(r)$ the
Bessel function, and $r=2\sqrt{3}t/J$. 

For our problem this ``path-independent energy'' approximation 
also assumes that one can neglect the renormalization of the
``bulk'' spin-polaron  wave-function due to the vicinity of the ADW. 
However, there is a concern that this
assumption may overestimate the energy of the excitation.
Near the ADW there is less energy required to create a spin flip,
therefore, there should be a subset of strings having a lower
energy than the string of the same length in the bulk. 
That is, there is a question if the modification of the spin-polaron
wave-function is  really negligible.
While we will demonstrate below the adequacy of the above
approximation, it  can be shown that one can 
consider the problem more rigorously  using the same approach by
taking into account the
energy of each string {\it exactly} up to a certain length $l_c$ and
applying the path-independent assumption only for $l>l_c$.
Technically, it means that one can find all $\omega_i$ in
Eq. (\ref{S}) up to some length of the string, use the explicit
continued-fraction form of $\Sigma(\omega)$ from Eq. (\ref{S}) up to
the same length $l_c$, and then use the approximate solution for the
continued-fraction 
$\Sigma(\omega)$ from Eq. (\ref{S1}) for $l>l_c$.
In other words, the modification of the spin-polaron wave-function in
the vicinity of the ADW can be consistently taken into account within
the same approach.

\begin{figure}[t]
\includegraphics[angle=270,width=3in,clip=true]{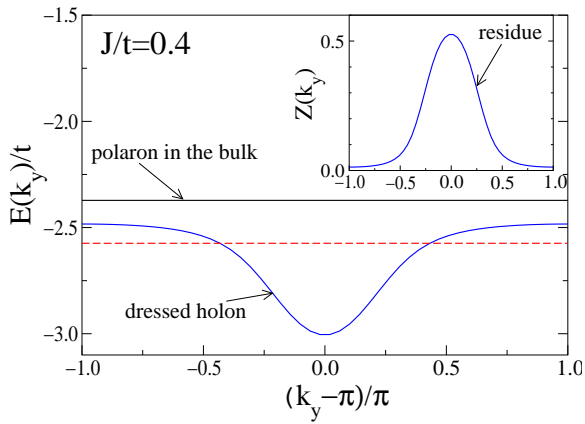}
\caption{Energy of an ADW elementary excitation $E(k_y)$ vs $k_y$
(solid curve). 
Solid straight line is the energy of the spin polaron in the bulk
$E^{sp}$. These
energies are relative to the energy of the {\it static} hole in a
corresponding magnetic background, that is, the {\it absolute} 
energy of the ADW excitation is $E_{wall}=(L_y-1)J/2$ higher than that
of the spin polaron.
Dashed line is the {\it total} 
energy per hole in the half-filled 1D stripe band $E_{1/2}$, energy of
the pure Ising state being subtracted. Since the reference energy for
$E_{1/2}$ and $E^{sp}$ is the same, they can be compared directly. 
Inset: residue of the Green's function versus $k_y$ (solid curves).}
\label{fig_9}
\end{figure}

The results of such calculations  for a representative value $J/t=0.4$
with $l_c=4$ 
are shown in Fig. \ref{fig_9}. The energy of the 
lowest pole of the Green's function 
versus $k_y$ is shown by the solid line. 
As in Fig. \ref{fig_6} there is an infinite off-set of the
renormalized holon-spin-polaron band from the spin-polaron energy by
the magnetic energy of the domain wall, $E_{wall}=(L_y-1)J/2$. Since the
effective band is significantly narrowed in comparison with the free
band the energy range shown in Fig. \ref{fig_9} is smaller than that
in Fig. \ref{fig_6}.
One can see that the energy of the
1D ADW excitation with the energy of the magnetic
background subtracted 
is now lower than the energy of the spin polaron in the bulk at
all $k_y$ which means that the ``dressed'' 1D band is
better for optimizing the kinetic energy than the homogeneous
spin-polaron state.

Another informative
quantity, the residue of the Green's function $Z(k_y)$,
is shown in Fig. \ref{fig_9} (inset). It gives a measure of the amount of
``bare'' holon in the wave function of the elementary excitations. 
We avoid the use of the ``quasiparticle residue'' since the holon is not a
quasiparticle in a strict sense, since its overlap with the physical
electron is zero.
However, the ``residue'' preserves its original meaning, which refers 
to the renormalization of some ``bare'' excitation.
One can see
that a significant part of the initial holon at $k_y=\pi$ resides inside
the wall (about one-half)
and that almost all its weight is transferred to strings at
$|k_y-\pi|>\pi/2$. Because the band is very flat at the same
$|k_y-\pi|>\pi/2$, 
the velocity of the elementary excitations is much slower than
the bare Fermi velocity $v_F^0=2t\sin k_F /\hbar$, in agreement with
experimental findings.\cite{Bal} 
It is easy to show that the velocity at the Fermi level for our 1D band
is $v_F=v_F^0 \, Z(k_F)$. 
Generally, both the flatness of the top of the band and vanishing
quasiparticle residue argue that the excitation at those $k$ are only
weakly attached to the stripe. The closeness of the
``bulk'' energy band to the top of the 1D stripe band 
suggests that the 1D excitations at the
top of its band are not too different from the ``bulk'' spin polarons 
in the homogeneous AF.

\subsection{Stripe energy}

\begin{figure}[t]
\includegraphics[angle=270,width=3in,clip=true]{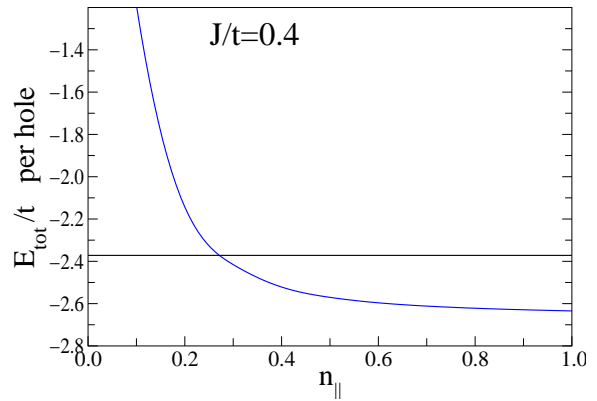}
\caption{Total energy of the system with ADW per hole
versus 1D hole density $n_{\parallel}$ (curve). 
Horizontal line is the total 
energy of free spin polarons in the homogeneous AF per hole. In both
cases the energy of the pure Ising state is subtracted, $J/t=0.4$.}
\label{fig_9a}
\end{figure}

For the  chosen value of  $J/t=0.4$  
the energy of the holon at the bottom of the band is lower than the
energy of the spin polaron by 
about $1.5J$, that is the energy of three ``wrong''  bonds. 
Then the rough estimation gives that the magnetic energy of the domain
wall will be 
compensated when in average every fourth site along the initial 1D
chain is taken by the hole. That is, the kinetic energy will make the
stripe to be the ground state at 1D linear hole concentration
$n_{\parallel}>n^c_{\parallel}\simeq 1/4$ (for $J/t=0.4$). A more accurate
consideration should
have the band-filling effects taken into account. Within the
rigid-band approximation the total energy of the system per hole with
the energy of the pure Ising state subtracted is
\begin{equation}
\label{E}
E_{tot}/N_h=\frac{J}{2}\left(\frac{1}{n_{\parallel}}-1\right)
+\frac{1}{2\pi n_{\parallel}}\int_{\pi-k_F}^{\pi+k_F}E(k_y)dk_y \ .
\end{equation}
where the first term is the domain wall 
energy and the  second term is the kinetic energy
of the free ``quasiparticles'' filling the effective 1D band up to $k_F=\pi 
n_{\parallel}$. The total energy per hole of this
1D band indeed crosses the spin-polaron energy at around
$n_{\parallel}\simeq 0.3$ as shown in Fig. \ref{fig_9a}.
The total energy per hole at half filling, $E_{1/2}\simeq -2.57t$, is shown
in Fig. \ref{fig_9} by the dashed line.
Above $n_{\parallel}\approx 0.5$ the energy (\ref{E}) versus 
$n_{\parallel}$ is almost constant \cite{CCNB} with the energy
difference between the stripe and no-stripe spin-polaron states of
about $-0.6J$.
This  behavior of $E_{tot}$ v.s. $n_{\parallel}$, the
absolute value of the energy difference, 
as well as the value of $n^c_{\parallel}$ are,
of course, functions of the $J/t$ ratio. 
However, these quantities only weakly
depend on $J/t$ in the realistic range $0.1<J/t<0.5$  with
$n^c_{\parallel}$ shifting towards 
zero for smaller $J/t$ 
(see also discussion of $E_{tot}$ v.s. $n_{\parallel}$ in
Sec. \ref{results}). 

An interesting question is what kind of energy scale defines the
difference between a homogeneous state with spin polarons and a stripe
state at some fixed $n_\parallel$ (say $n_\parallel=1/2$), $\Delta
E=E^{sp}-E^{ADW}_{1/2}$. By letting $J/t\rightarrow 0$ we observe that
$\Delta E\rightarrow 0$ as well, and thus it cannot be proportional to
$t$. This is in agreement with the expectation that at $J=0$ all
magnetic configurations should become degenerate.
It is nevertheless hard to tell analytically or numerically 
if the leading term in $\Delta E$ scales with $J^{2/3}$ 
(however, see the discussion at the end of the section).

\subsection{Densities, electron distribution function}

The knowledge of the single-particle Green's function not only allows
us to calculate the energetic stability  of the stripe as a function of
$n_{\parallel}$ and $J/t$\cite{CCNB} 
but also enables us to find spatial profiles of
the charge and spin densities, $N(x)$ and $|\langle S^z\rangle (x) |$, as
well as the
electron distribution function $n_{\bf k}$ for the stripe state
\cite{remark}.
To evaluate those quantities one needs to know  the wave function of
the 1D band excitation written in terms of the operators of spins and
holes. 
If such a wave function is known for each $k_y$ in the effective band
then the distribution of the hole density and the ``damage'' of 
the spin density from the hole and ``strings''  can be extracted.
Within the rigid-band picture these wave functions are orthogonal at 
different $k_y$ and thus the density profiles are given by the 
superposition of the contributions from each $k_y$:
\begin{eqnarray}
\label{N_x}
&&N(x)=\sum_{k_y<k_F} n^h(x,k_y) \ , \nonumber\\
&&|\langle S^z\rangle (x)| =\sum_{k_y<k_F} |\langle s^z\rangle (x,k_y)| 
\ ,
\end{eqnarray}
where
\begin{eqnarray}
\label{n_x}
&&n^h(x,k_y)=\langle \tilde{k}_y| \tilde{c}_j\tilde{c}_j^\dag
|\tilde{k}_y\rangle, 
\nonumber\\
&&|\langle s^z\rangle (x,k_y)| =\langle \tilde{k}_y| \hat{S}^z_j
|\tilde{k}_y\rangle\ ,
\end{eqnarray}
where $j=(x,y)$ and the choice of the $y$-coordinate is arbitrary.
\begin{figure}[b]
\includegraphics[width=3in,clip=true]{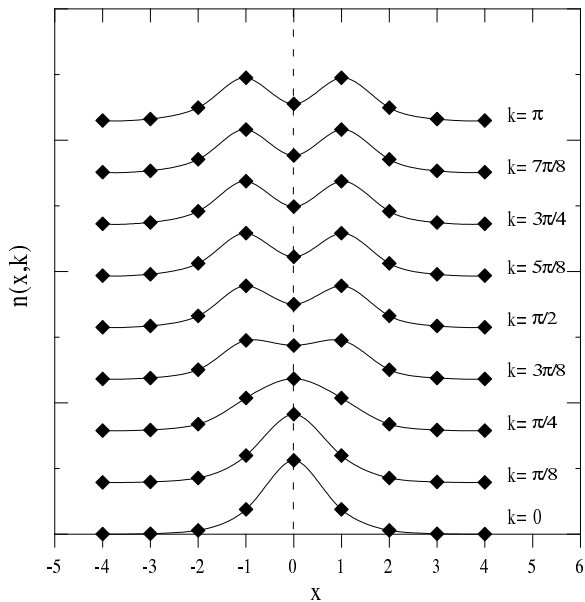}
\caption{The hole density modulation
within the single-hole $k_y$-eigenstate  for different momenta $k_y$,
$k=|k_y-\pi|$.
One can see that at the momenta away from the bottom of the band
the hole is mostly spread around  the stripe.}
\label{fig_9b}
\end{figure}

Such a wave function for our composite excitation can be written in a 
variety of ways.
The simplest one is to write it as a linear combination of the 
components with 
strings of different length \cite{SS,Eder1,Ed1}:
\begin{eqnarray}
\label{wf}
|\tilde{k}_y\rangle=\sum_{l=0}^{\infty} C_{k_y,l}|l,k_y\rangle \ ,
\end{eqnarray}
where $|0,k_y\rangle$ is a pure holonic state and under the index $l$ 
we also understand a summation over different 
paths of the same length $l$.
Then, using a retraceable-path approximation, it is easy to see that 
for each specific path
\begin{eqnarray}
\label{C_wf}
C_{k_y,l}=\frac{t\, \sqrt{Z_{k_y}}}{E_{k_y}-\omega_1-\dots -\omega_l-
\Sigma_{l+1}(E_{k_y})}\ ,
\end{eqnarray}
where multiplier $\sqrt{Z_{k_y}}$ comes from the normalization 
condition
$\langle\tilde{k}_y|\tilde{k}_y\rangle=1$.
An equivalent expression for the spin-polaron case has been obtained
in Refs.~[\onlinecite{Reiter,Oleg}] using the 
spinless-fermion-Schwinger-boson
representation for the original constrained fermion operators.

Since the hole density and the spin density are given by the 
averages of the 
local (on-site) 
operators their calculation using Eq. (\ref{wf}) is quite 
straightforward because they
are diagonal in the string basis and do not provide ``transitions'' 
between  
different components of the wave function (\ref{wf}). 
Calculation of $n_{\bf k}$ 
is technically
more cumbersome since it is given by the non-local averages. One can 
always 
rewrite $n_{\bf k}$ as
\begin{eqnarray}
\label{n_k}
n_{\bf k}=\langle{\tilde{c}}^\dag_{\bf k}{\tilde{c}}_{\bf k}\rangle
=\frac{1}{N}\sum_i\big[
\langle{\tilde{c}}^\dag_i{\tilde{c}}_i\rangle +\sum_{{\bf d}\neq 0}
e^{i{\bf k}{\bf d}}\langle{\tilde{c}}^\dag_i{\tilde{c}}_{i+d}\rangle\big]\ 
,
\end{eqnarray}
where we drop the spin index for clarity. One can see that 
the matrix
 elements between different ``strings'' are essential.
For the details of calculations of $n_{\bf k}$ in the $t$-$J$ model 
we refer to the earlier works, Refs.~[\onlinecite{CLG,Eder2,Prelovsek_ARPES}].

As an example of the calculation of the discussed quantities 
we show our results for $n^h(x,k_y)$, Eq. (\ref{n_x}), in
Fig. \ref{fig_9b} for $J/t=0.4$.
This quantity shows the hole density modulation in the direction
perpendicular to the stripe
within the $k_y$-eigenstate in the 1D band 
for different momenta $k_y$. 
One can see that at the momenta away from the bottom of the band 
the hole is mostly spread around  the stripe.
Further results on these and other quantities are given in the context 
of the comparison with the numerical data, Sec. \ref{results}.

\subsection{Other questions}

\begin{figure}[t]
\includegraphics[angle=270,width=3in,clip=true]{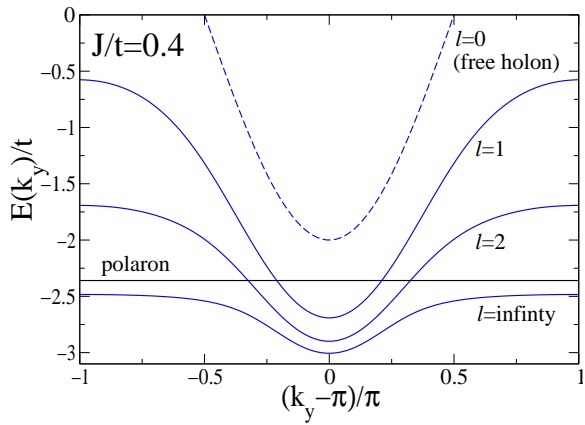}
\caption{Energy of an ADW elementary excitation $E(k_y)$ v.s. $k_y$ for
``strings'' with $l=0, 1, 2,
\dots\infty$ included in the wave function.}
\label{fig_12}
\end{figure}

There are few other issues we would like to address
here. First, if one is to start from the strictly 1D stripe and is to 
gradually ``relax'' the initial restriction
for the hole to be kept in such a  1D chain,
 how soon does one reach the ``true''
eigenvalue of the $t$-$J_z$ Hamiltonian? 
That is, let us allow the ``bare'' holon to have an 
admixture of strings of length $l=1,2,\dots$ and 
see when the results converge and become independent of $l$. 
Figure \ref{fig_12} shows the
energy bands for such ``restricted-string-length''
problems. $l=0$ corresponds to a free holon,
Fig. \ref{fig_6}, $l=1$ and $l=2$ are for the cases
where the ``strings'' of length 1 and 2 are
included in the approximation. $l=\infty$ is our answer
from Eqs. (\ref{G})-(\ref{S1}), Fig. \ref{fig_9}.
\begin{figure}[b]
\includegraphics[angle=270,width=3in,clip=true]{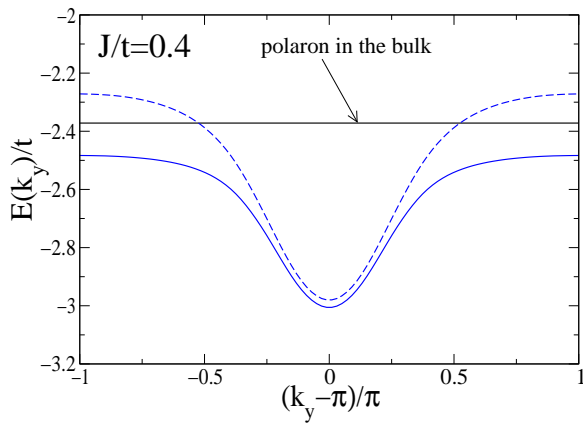}
\caption{1D energy band for the composite excitation. Renormalization
of the spin-polaron propagator at the vicinity of the ADW is included
(solid line), not included (dashed line).}
\label{fig_13}
\end{figure}

This comparison demonstrates one more time that the ``bare'', or
almost
``bare'' 1D excitation with some admixture of the
``transverse fluctuations'' is certainly not a good 
approximation for the description of the stripe charge
excitation. It is especially true for the finite
concentrations where large momenta around
$k_F^{1D}\simeq\pi/2$ are most important.

As an alternative starting approximation we argue,
once again, for the 
picture of a holon strongly coupled to the ``bulk''
spin-polaron excitation, Fig. \ref{fig_8}.
The dispersion obtained without renormalization
of the spin-polaron propagator in Fig. \ref{fig_8}(b) 
is shown by the dashed line in Fig. \ref{fig_13}, 
which is already very close to the result where such a
renormalization is taken into account (solid line, from 
Fig. \ref{fig_9}) \cite{remark1}.

The second issue is the following. Since there are two
types of kinks in the system (kink and anti-kink) 
which can be associated with the pure 
holonic excitation, Fig. \ref{fig_3}, one can have two 
different ``species'' of excitations having 
opposite ``geometrical'' quantum numbers
\cite{Oleg_T}.
One may then think of them as essentially independent
particles populating (quasi-) independent
bands. 

A little thinking brings such a logic into a paradox:
the case when the chain is completely filled (all
sites are occupied by holes) in this language will
correspond to two half-filled (conducting?) 
bands of different ``flavor''. One resolves the
paradox 
by noting that these ``extra'' quantum numbers 
should not lead to the increase of the total number 
of states in the system.
In our case there cannot be more than one hole at any
site while the assumption of ``independence''  of
opposite species implicitly doubles the Hilbert space.
Therefore, if one wants to keep two ``flavors'' in
the problem, an infinite on-site repulsion between
them 
should be introduced. That latter problem in one dimension maps
exactly onto the problem of spinless, single-flavor 
particles \cite{1D_mapping} 
and thus the single-band approach used throughout 
this work is justified.
Note that in the presence of the 2D degrees
of freedom where the stripe as a whole can make wiggles
the problem of these ``geometrical''
quantum 
numbers becomes more complicated \cite{Oleg_T}.

The last question is what energy scale defines 
the ``free'' kinetic energy difference  between the stripe and 
no-stripe states. Naively, such $\Delta E$ should be governed by the 
kinetic $t$-term since the hole
motion can be made free in a 1D structure.
However, the energy of the order $\sim JL$ is paid to ``prepare'' such
a structure in a 2D AF.
Let us consider the $t\gg J$ limit for the 1D hole motion
and assume that the length of the ADW is a free parameter
which would minimize the total energy. 
In the continuum limit $\langle E_{kin}\rangle =
-2t+A t/L^2$, $\langle E_J\rangle\sim JL$, and the minimum
of the energy is achieved at $L_{optimal}\sim
(J/t)^{1/3}$. The corresponding energy is  $E_{min}\simeq
-2t+\alpha (J^2t)^{1/3}$. One recalls an almost
identical consideration of the ``retraceable-path'' motion of the hole
by the ``strings'' in the spin-polaron problem 
which also gives $\langle L_{string}\rangle \sim 
(J/t)^{1/3}$ and $E_{sp}\simeq 2\sqrt{3}t +\beta
(J^2t)^{1/3}$, see Ref.~[\onlinecite{BNK}]. 
Therefore, there is no new energy scale, different from the
homogeneous N\'eel problem, introduced by the domain wall 
and the ``prepared-path'' motion in 1D ADW is, in fact, not too
different from the ``retraceable-path'' motion of the hole within 
the spin polaron. 
One may conclude that the same scale $\sim (J^2 t)^{1/3}$ should
govern the energetic balance favoring the stripe.\cite{remark2} 
This is yet another argument  that the stripes
are the outcome of the same tendencies which are seen already for the
single-hole problem.

\section{Numerical approach}
\label{Numerics}

The numerical study of the $t$-$J_z$ model utilizes the DMRG method
for the clusters up to 
$11\times8$ with various boundary conditions (BC's). 
The BC's at the left and right sides are always
open; if the BC's at the top and bottom are periodic we refer to
the BC's as cylindrical.
A staggered magnetic field, typically of size 0.1$t$,
can be applied at the open ends 
in order to enforce the ADW inside the system, or enforce it to stay
homogeneous. Typically we keep 1000-2000 states per block, and
perform up to a dozen finite system sweeps. Typically the
truncation error was $5\times10^{-5}$.

\begin{figure}
\includegraphics[width=3in,clip=true]{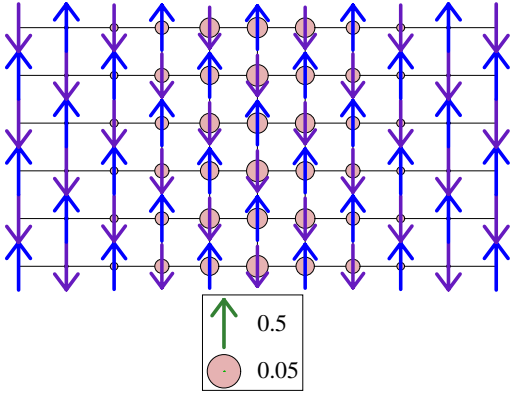}
\caption{$11\times 6$ cluster with one hole, cylindrical BC's,
and $J/t=0.35$. 
The size of the circles is proportional to the on-site hole
density.}
\label{fig_14}
\end{figure}

\begin{figure}
\includegraphics[width=3in,clip=true]{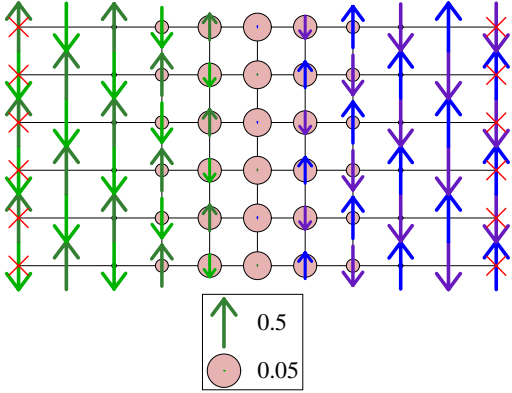}
\caption{$11\times 6$ cluster with one hole, 
field applied at open ends (crosses) 
to stabilize ADW, $J/t=0.35$.}
\label{fig_15}
\end{figure}

\begin{figure}
\includegraphics[width=3in,clip=true]{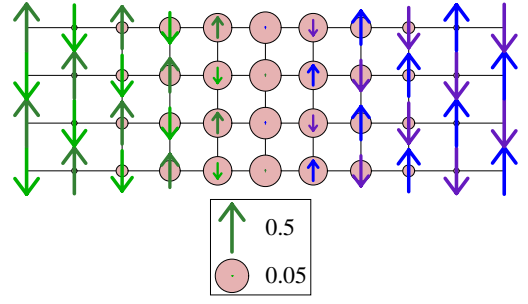}
\caption{$11\times 4$ cluster, one hole, open BC's in the narrow
direction, no staggered field, cylindrical BC's,
$J/t=0.2$. The ground state here contains a stripe.}
\label{fig_16}
\end{figure}

\begin{figure}
\includegraphics[width=3in,clip=true]{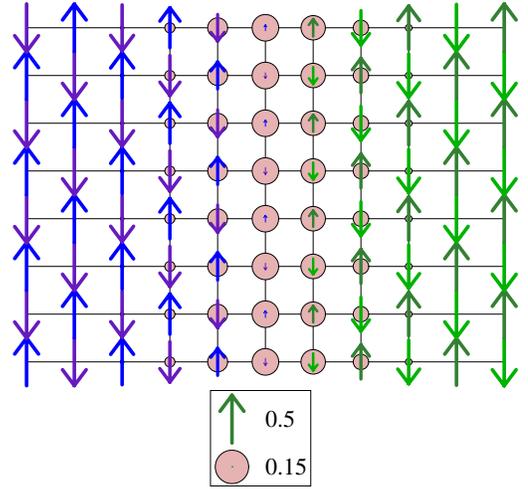}
\caption{$11\times 8$ cluster with four holes, cylindrical BC's,
and $J/t=0.35$. The ground state contains a stripe.}
\label{fig_17}
\end{figure}

Figures \ref{fig_14}-\ref{fig_17} show several examples of the
states within the different clusters. Figure \ref{fig_14} shows the 
$11\times 6$ cluster with one hole, cylindrical BC's, and with
$J/t=0.35$. The ground state is
homogeneous in this case. 
Figure \ref{fig_15} shows the 
same $11\times 6$ cluster with one hole, 
but with a staggered field applied at the open ends, $J/t=0.35$. 
The ground state is an ADW with the hole bound to it.
Figure \ref{fig_16} shows the narrower
$11\times 4$ cluster with one hole, no staggered field,
cylindrical BC's, and at smaller $J/t=0.2$. 
The system spontaneously forms a stripe in the ground state for this
system. One may conclude from here that in the $11\times L_y$ system 
with the number of holes $N_h\simeq L_y/4$ one will have a stripe in
the ground state for $J/t\alt 0.2$.
Figure \ref{fig_17} shows  the largest
$11\times 8$ cluster with four holes, no staggered field,
cylindrical BC's, and $J/t=0.35$. In this case the ADW is also
spontaneously formed with the holes occupying the border between the
antiferromagnetic domains. 

\section{Results}
\label{results}
\subsection{Single-hole problem}

\begin{figure*}
\includegraphics[width=1.6in]{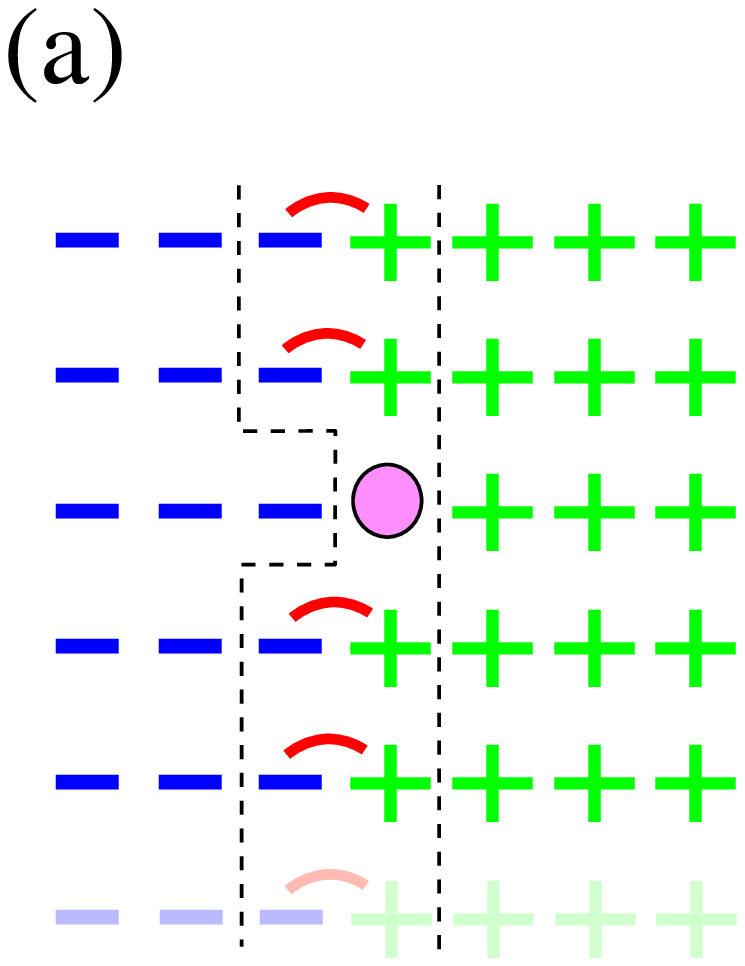} \hskip 1.5cm
\includegraphics[width=1.6in]{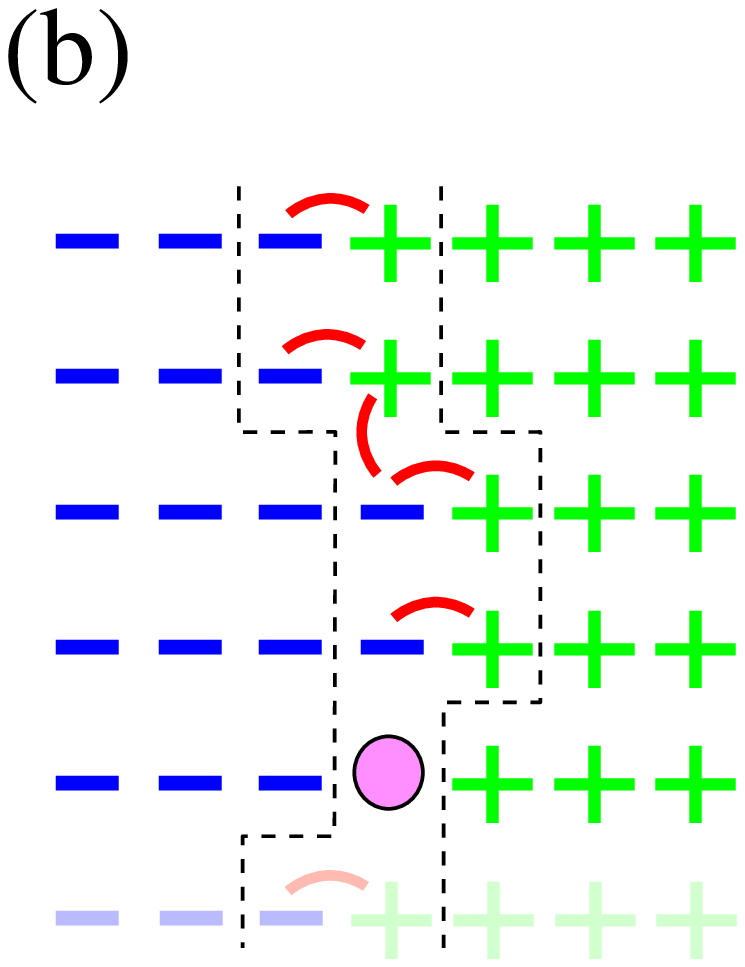} \hskip 1.5cm
\includegraphics[width=1.6in]{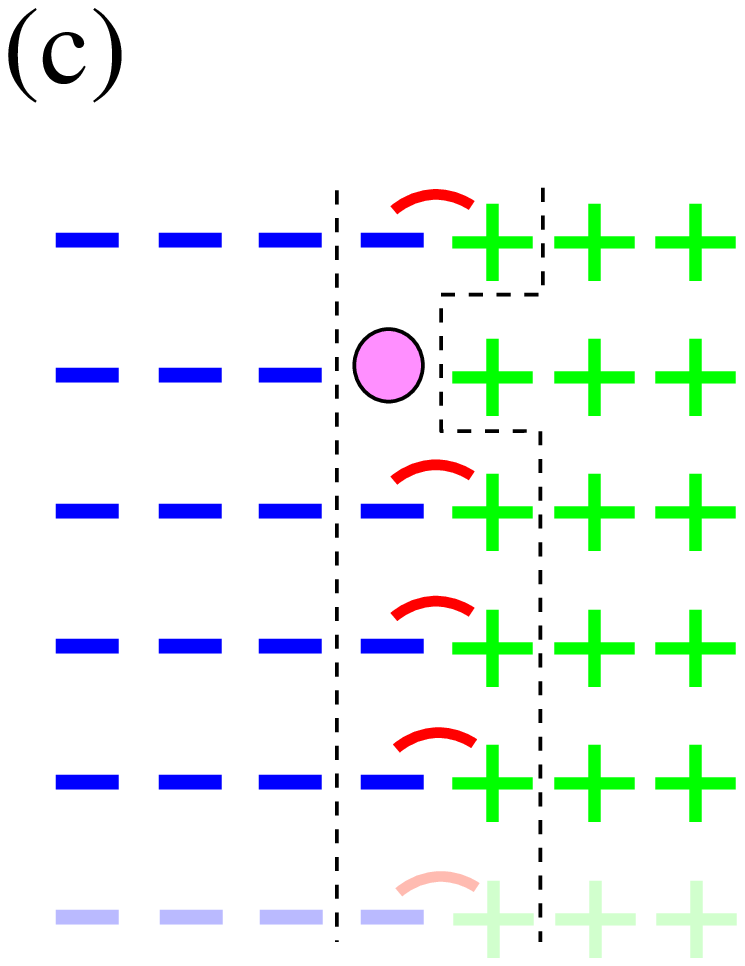}
\caption{A schematic picture of a $7\times 5$ cluster with periodic BC's and
ADW. (a) One hole, straight ADW is enforced by periodic BC's, (b) holon and
spinon in the straight ADW, (c) spinon and holon recombined after the
circular motion of the hole around the cylinder, the ADW is translated
in $x$-direction as a result. The shaded row at the bottom is
equivalent to the top row and is shown to emphasize periodic BC's in
$y$-direction.}
\label{fig_18}
\end{figure*}

The first problem we would like to clarify is the single-hole
excitation problem in the stripe configuration. We therefore
consider different clusters with an ADW induced by
staggered fields for various values of $J/t$. 
\subsubsection{Boundary conditions}
\begin{figure}[b]
\includegraphics[width=2.5in,clip=true]{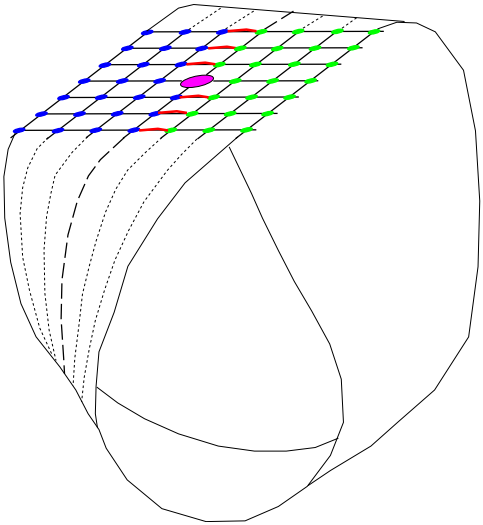}
\caption{A schematic view of $7\times 7$ cluster with M\"obius BC's. An
ADW in the center of the cluster and a ``wiggle'' are induced by M\"obius BC's.}
\label{fig_19}
\end{figure}

We would like to remark here that simple 
cylindrical or periodic BC's do not quite work for
the purpose of the study of the single excitation for the
following reason. As we have discussed in Sec. \ref{Analytics} the
``core'' of the composite excitation is a holon, that is, an excitation
carrying the kink, Fig. \ref{fig_5}, while periodic BC's require the
domain wall to be closed on itself around the cylinder, Fig.
\ref{fig_18}(a). This means that ({\it i}) there will be 
two excitations in the wall, holon
and spinon, instead of one, Fig. \ref{fig_18}(b), ({\it ii}) the free holon
movement is frustrated by the spinon, ({\it iii}) periodic BC's induce an
artificial (absent in $L_y\rightarrow\infty$ limit) stripe meandering. The
meandering is induced by the process of holon recombination with the
spinon with the subsequent change of the side of the domain wall. Then,
when the holon completes the full circle around the cylinder the whole
domain wall is translated in the $x$-direction, Fig. \ref{fig_18}(c).

The way to avoid such a frustration of the ``free'' holonic motion in the
finite system is intuitively evident \cite{remark3}. Since the holon
carries a topological (``geometrical'') charge the topological BC's are to
be used. In our case it is M\"obius BC's, Fig. \ref{fig_19}. 
One can see that with these BC's the holon motion is unfrustrated.
The number of sites should be odd to avoid the
frustration in the spin system. For the case of the M\"obius BC we,
therefore, study 
clusters $11\times 7$, $11\times 5$, and $11\times 3$, and in the case of
cylindrical BC's,
the clusters are $11\times 8$, $11\times 6$, and $11\times 4$.
\subsubsection{Energy}

\begin{figure}[t]
\includegraphics[angle=270,width=3in,clip=true]{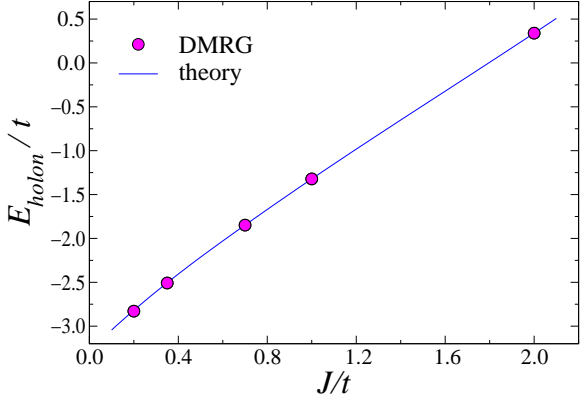}
\caption{$J/t$ dependence of the ground-state energy
of: the single hole in $11\times 7$ cluster with M\"obius BC 
(DMRG, circles) and
the single holon-spin-polaron in the infinite stripe (theory,
solid line).}
\label{fig_20}
\end{figure}

Figure \ref{fig_20} shows the 
 $J/t$ dependence of the ground-state energy
of a single hole in the $11\times 7$ cluster with a M\"obius BC, energy 
of the undoped system being subtracted. 
The theoretical curve shows the $J/t$ dependence of the 
ground-state energy
of the single holon-spin-polaron, that is, the energy
of the bottom of the effective 1D band in Fig. \ref{fig_9},
obtained from Eqs.~(\ref{G})-(\ref{S1}).
Note that the reference energy in Sec. \ref{Analytics} differs from that
used here by the constant $2J$, the quantity associated with the energy
of four bonds ``broken'' by the hole.
We would like to note that there is no energy adjustment between
numerical and analytical data used in Fig.
 \ref{fig_20} and it is not the result of the best fit.
The maximal discrepancy of numerical and analytical results 
in Fig. \ref{fig_20} is about $\sim 0.1\%$. 
The closeness of agreement is even
better than in the case of similar calculations 
for the spin polaron compared to the Exact
Diagonalization data \cite{CP}. The reason for that has been discussed in
Sec. \ref{Analytics}. The Trugman paths, which are beyond the retraceable
path approximation, are the source of the discrepancies in the case of the
spin polaron. In our case the Trugman paths do not exist in their original
form. 
\begin{figure}[b]
\includegraphics[angle=270,width=3in,clip=true]{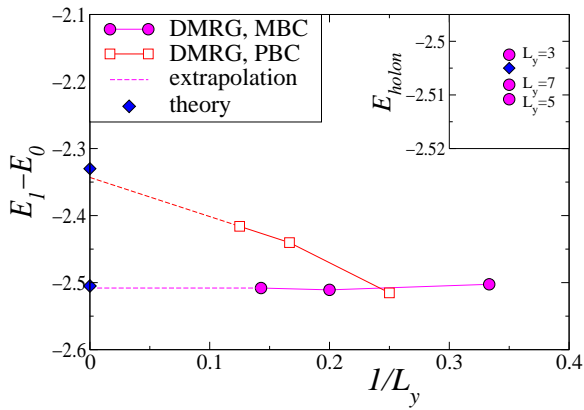}
\caption{The ground-state energy v.s. inverse linear size of the
cluster. M\"obius BC's (circles) correspond to the case of a single holon,
periodic BC's 
(squares) induce the state with the holon and spinon. Theoretical results
for the energy of the free holon and holon $+$ spinon are put on the
$y$-axis (diamonds). Lines are guides to the eye. 
Inset shows M\"obius BC's and theoretical data in a strongly magnified
scale. $J/t=0.35$.
}
\label{fig_21}
\end{figure}

One can also study the ground-state energy dependence on the size of the
system. In Fig. \ref{fig_21} we show $E_{GS}$ versus $1/L_y$ for
$L_y=3,5,7$ (M\"obius BC's) and for $L_y=4,6,8$ (periodic BC's) together
with the theoretical results for the infinite system, $J/t=0.35$. 
For the case of M\"obius BC's the theoretical point at $1/L_y=0$ is 
 $E_{GS}=E_{holon}$, for the case of periodic BC's 
the theoretical point represents the sum of the holon and
spinon energy, $E_{GS}=E_{holon} +J/2$. That is, we assume that they do
not form a bound state in the thermodynamic limit. One can see essentially
negligible finite-size effects on the energy of the ``free''
holon-spin-polaron (M\"obius BC's, also see inset). 
This feature can be anticipated since the
excitation is a band-like state and the ground-state wave-vector
belongs to the 
reciprocal space of all three clusters. The linear extrapolation to
$1/L_y=0$ in
Fig. \ref{fig_21} was made from $L_y=5,7$ and $L_y=6,8$ for M\"obius
BC and periodic BC 
results, respectively.
\subsubsection{Density}
\begin{figure}[t]
\includegraphics[angle=270,width=3in,clip=true]{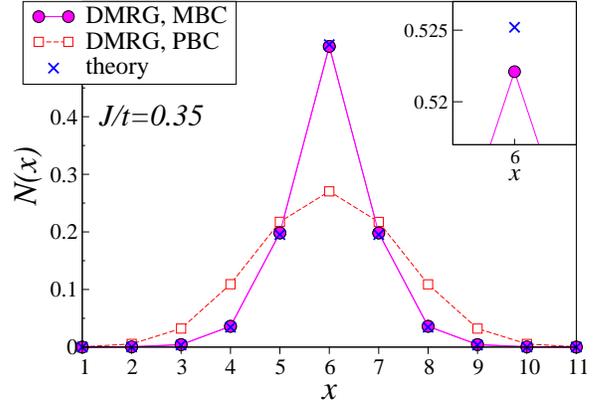}
\caption{Hole density distribution across the stripe at $J/t=0.35$.
Data for M\"obius BC's, $11\times 7$ cluster (circles), periodic BC's
$11\times 8$ 
cluster, and theoretical results (crosses) are shown.
Lines are guides to the eye. Inset shows M\"obius BC's and
theoretical data for $x=6$ in a strongly magnified scale. Enlarged
area is shown by the box.}
\label{fig_22}
\end{figure}

Figure \ref{fig_22} shows the results for the hole density distribution
across the stripe for the ground state of the single hole at
$J/t=0.35$. DMRG data from the $11\times 7$ cluster with M\"obius BC's 
are shown by the circles. Theoretical results 
obtained from Eqs.~(\ref{n_x})-(\ref{C_wf}) are shown by crosses.
The total disagreement of the
numerical M\"obius BC data and analytical result  is $\sum_x
|N_{num}-N_{th}|/N_{num}\simeq 0.3\%$. The same Fig. \ref{fig_22} 
shows the density profile of the hole distribution along the $x$-axis for the
case of periodic BC's.
We note that the hole density modulation in the case of periodic BC's is
essentially induced by the repulsion of the open boundaries in the
$x$-direction. If the system would be significantly
wider in the $x$-direction the
circular meandering effect, which induces the transverse motion of the
domain wall, would spread the density homogeneously. The same would be 
true in
the case of periodic BC's in the $x$-direction. This just demonstrates the
following paradox: a finite system with periodic BC's in all directions
would show a homogeneous ground state, as in Ref.~[\onlinecite{HM}],
whose wave 
function may be, in fact, given by the superposition of
slowly meandering stripes. In that case the stripe should
be detected not from the density profile but from the instantaneous
spin-spin correlation function, which should exhibit a strong anti-phase
component.
\begin{figure}[t]
\includegraphics[angle=270,width=3in,clip=true]{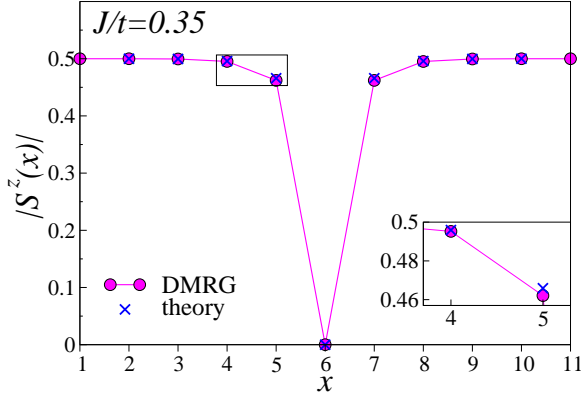}
\caption{The spin-density $|\langle S^z\rangle (x)|$ v.s. $x$
at $J/t=0.35$. Data for M\"obius BC's, $11\times 7$ cluster (circles),
and theoretical results (crosses) are shown.
Lines are guides to the eye. Inset shows M\"obius BC's and
theoretical data for $x=4$ and $5$ 
in a magnified scale. Enlarged area is shown by the box.}
\label{fig_23}
\end{figure}
\begin{table}[b]
\caption{$\sum_i
\langle\tilde{c}^\dag_i\tilde{c}_{i+d}\rangle$ for several
coordination vectors ${\bf d}$. $J/t=0.35$.}
\vskip 0.5cm
\begin{tabular}{c|cc}
\ \hskip 0.5cm \ \ ${\bf d}$\ \hskip 0.5cm \ \ & 
\ \ \ \hskip 0.5cm DMRG \hskip 0.5cm
 \  \  \ & \ \ \ \hskip 0.5cm theory \hskip 0.5cm \ \ \ 
\\ \\
\hline \\
(1,0) & 0.1041 & 0.102 \\
(0,1) & 0.1216 & 0.115  \\
(1,1) & -0.1266 & -0.093  \\
(2,0) & -0.0030  & $\sim 0$  \\
(0,2) & -0.0041 & $\sim 0$ \\ \\ 
\end{tabular}
\end{table}

The density profile study of the single-hole problem is completed by the
$|\langle S^z\rangle (x)|$ data shown in Fig. \ref{fig_23}. Note that the
spin-density profile is not straightforwardly related to the hole density
since there is also a large contribution from the hole-induced strings of
misaligned spins affecting the average on-site spin values. For a single hole
in the stripe of the length $L_y$ 
the suppression of the spin-density should be
proportional to $1/L_y$ ($L_y=7$ in the case of the $11\times 7$
cluster). 
This is valid for all $x$
except the ``center'' of the stripe where the
motion is holonic. Since the holon is a topological excitation it 
borders two 1D domains with opposite staggered spin value.
Because of that any site along $x=x_0$ ($x_0=6$ in Fig.
\ref{fig_23}) always has average spin zero and this is not a $1/L_y$ effect.
The deviation of the theoretical results from DMRG for 
$|\langle S^z\rangle (x)|$ is more observable than for $N(x)$. This is
because of the subset of the paths which correspond to the processes of
hole departing and then rejoining the stripe in the other leg of the
ladder. Such paths can produce deviations from the $1/L_y$ character of the
spin-density modulation for $x\neq x_0$, similar to that at $x=x_0$,
which are hard to account for analytically. In
any case, the largest discrepancy between the theory and numerical
data for $|\delta S^z(x=x_0\pm 1)/S^z(x)|$ does not exceed $10\%$.
\subsubsection{Electron distribution function}
\begin{figure}[t]
\includegraphics[angle=270,width=3in,clip=true]{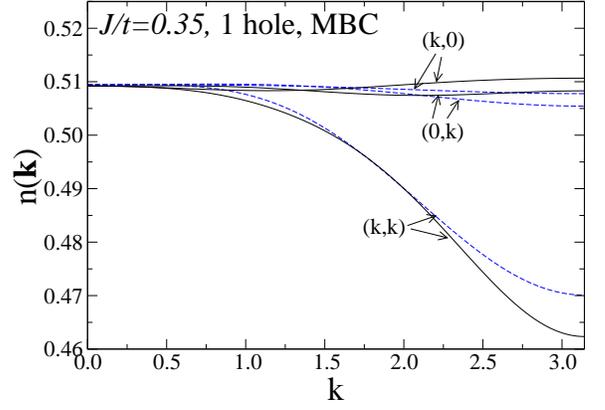}
\caption{$n_{\bf k}$ for one hole in M\"obius BC $11\times 7$ cluster for 
$(1,0)$, $(0,1)$,
and $(1,1)$ directions, $J/t=0.35$. Solid and dashed lines are DMRG
and theory results, respectively.}
\label{fig_24}
\end{figure}
\begin{figure}[b]
\includegraphics[width=3in,clip=true]{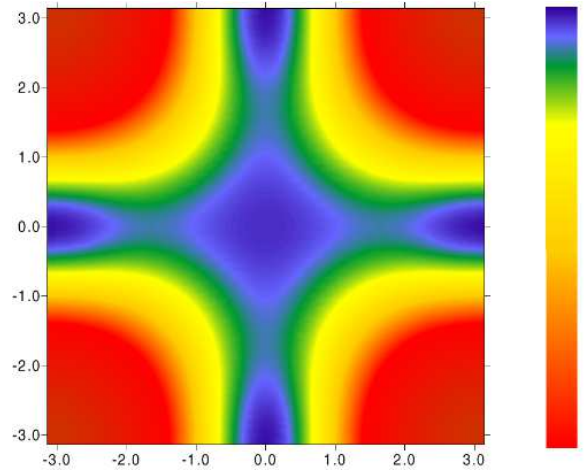}
\caption{2D intensity plot of $\bar{n}_{\bf
k}=[n(k_x,k_y)+n(k_y,k_x)]/2$ from DMRG for a single hole in M\"obius BC
$11\times 7$ cluster, $J/t=0.35$.}
\label{fig_25}
\end{figure}
\begin{figure*}
\includegraphics[width=1.5in]{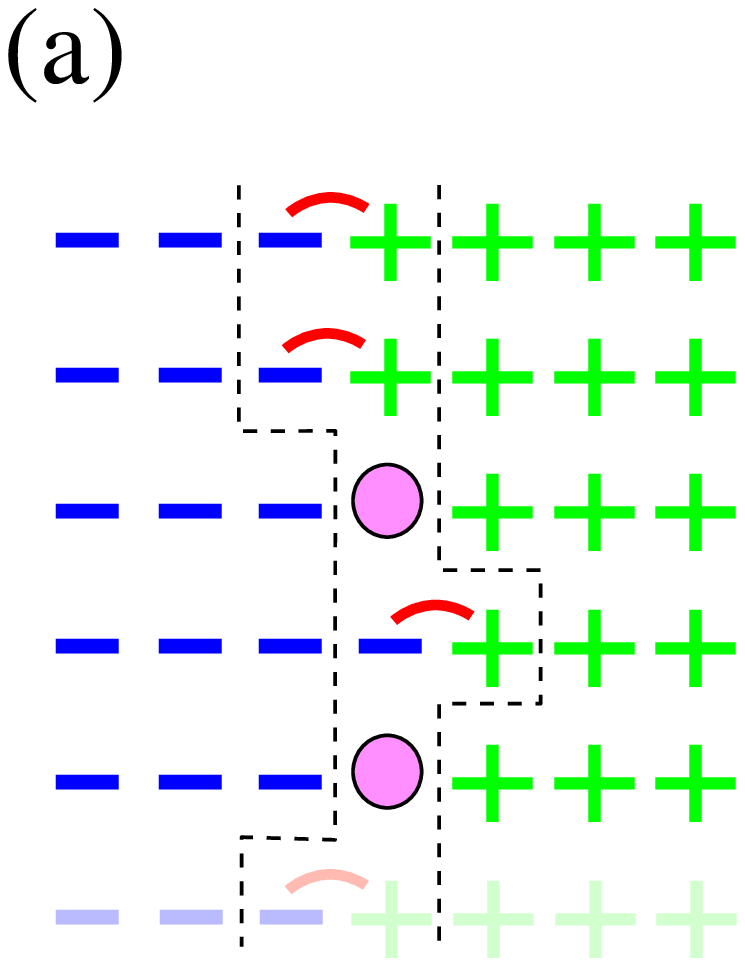}\hskip 0.5cm
\includegraphics[width=1.5in]{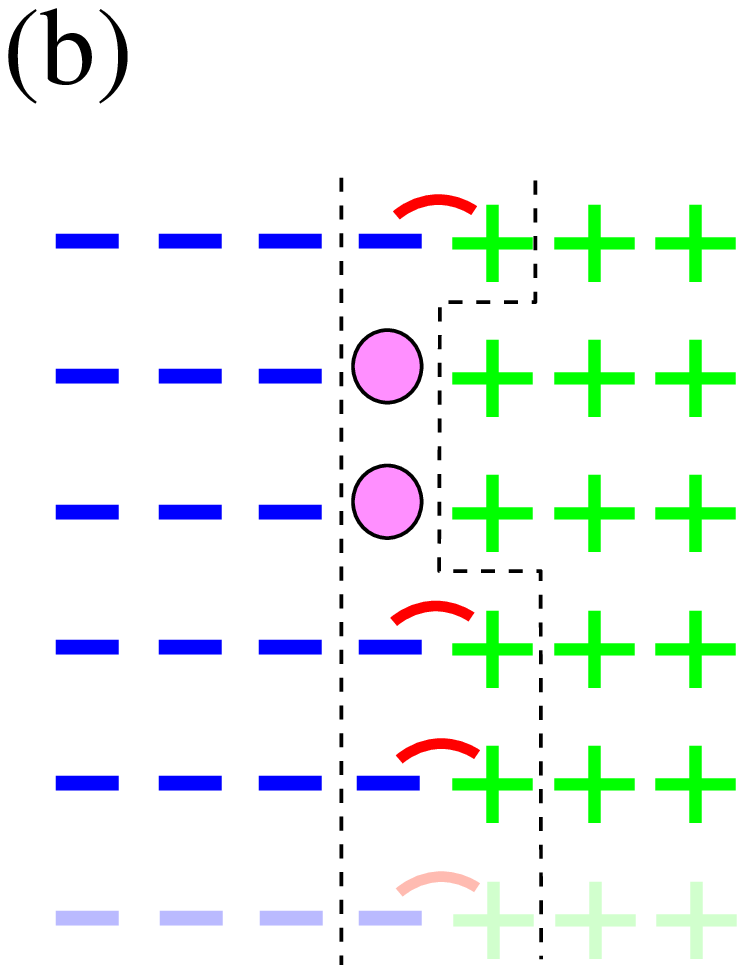}\hskip 0.5cm
\includegraphics[width=1.5in]{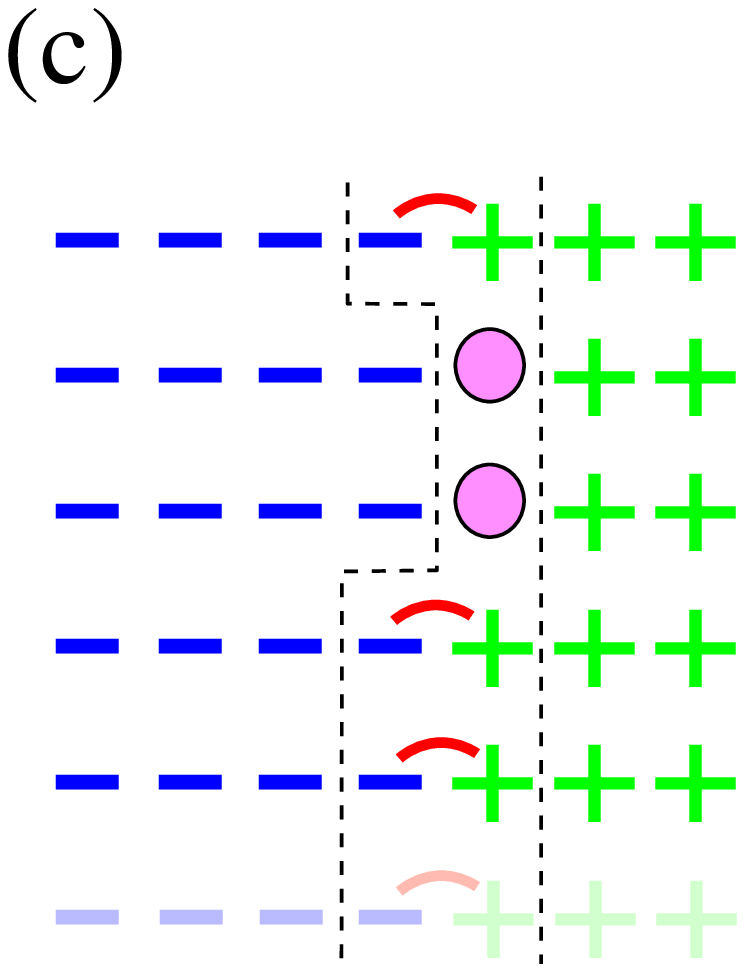}\hskip 0.5cm
\includegraphics[width=1.5in]{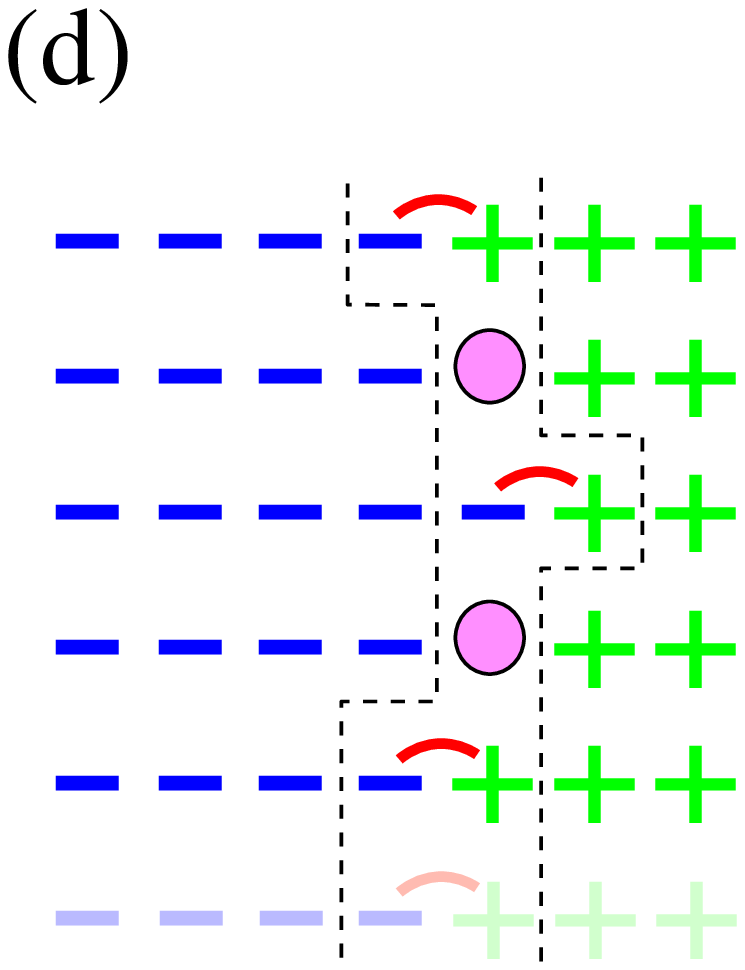}
\caption{A schematic picture of $7\times 5$ cluster with two holes in
an ADW. (a)-(d) hole motion in the periodic direction combined with
the transverse motion of both holes results in the 
translation of ADW in  $x$-direction. 
The shaded row at the bottom is
equivalent to the top row and is shown to emphasize periodic BC's in
$y$-direction.}
\label{fig_26}
\end{figure*}

The electron distribution function is compared as well. For this
quantity the 
analytical calculations are quite cumbersome and were restricted by
$|{\bf d}|=2$ in Eq. (\ref{n_k}). One can see from Eq. (\ref{n_k})
that, apart from the constant, $n_{\bf k}$ is given by the set of $\cos
({\bf dk})$ with the coefficients given by the $\sum_i
\langle\tilde{c}^\dag_i\tilde{c}_{i+d}\rangle$ averages. We list such
coefficients 
for several ${\bf d}$'s in Table I, where one can see a very good 
agreement of DMRG with the theory.

Figure \ref{fig_24} shows a plot of $n_{\bf k}$ in the $(1,0)$, $(0,1)$,
and $(1,1)$ directions. A more informative 2D intensity plot in Fig.
\ref{fig_25} shows the DMRG results for the electron distribution
function averaged over the $x$ and $y$ directions, $\bar{n}_{\bf
k}=[n(k_x,k_y)+n(k_y,k_x)]/2$. Two features are worth
discussing.

First, there is almost no anisotropy between 
the $k_x$ and $k_y$ directions in $n_{\bf k}$ for the highly
anisotropic stripe 
configuration, Fig. \ref{fig_24}. 
That can be, broadly speaking, interpreted as yet
another demonstration of the equal importance of the ``transverse'' and
``longitudinal'' kinetic energies \cite{Steve_Kampf}. 
The broad features in $n_{\bf k}$ in the $t$-$J$-like
models are understood as
coming from the ``fast'' (incoherent) motion of the hole inside the
quasiparticle, while the coherent part should show itself as a
$\delta$-peak at ${\bf k}={\bf k}_0$ proportional to the quasiparticle
residue  \cite{Eder2,CLG,Prelovsek_ARPES}. In the stripe case,
excitations are  not quasiparticles 
in a standard sense and, therefore, no sharp
features are to be expected. In fact, it is easy to show that  $n_{\bf
k}$ for a holon in a periodic 1D Ising chain
 is equal to:
\begin{eqnarray}
\label{n_k_1D}
n_{\bf k}=\langle k_0|\tilde{c}^\dag_{\bf k}\tilde{c}_{\bf k}|k_0\rangle
=\frac{1}{2}-\frac{1}{L} \cos k \cos k_0\ ,
\end{eqnarray}
where $k_0$ is the momentum of the holon. This is the simple consequence
of the fact that the holon is a zero-dimensional domain wall and that
$\langle\tilde{c}^\dag_i\tilde{c}_{i+d}\rangle$ averages 
for distances larger than
$d=1$ are identically zero.

Second, the general behavior of $n_{\bf k}$ in our problem
is similar to that for the 
spin polaron: it has a maximum at ${\bf k}=(0,0)$ and a minimum at
${\bf k}=(\pi,\pi)$ which is a simple consequence of the kinetic
energy minimization in the ground state\cite{Eder2}. 
However, its shape is cross-like
 rather than diamond-like (for the shape of spin-polaron $n_{\bf
k}$ see Refs.~[\onlinecite{Eder2,Prelovsek_ARPES}]). For spin polarons
$n_{\bf k}$ is 
mainly given  by the powers of $\gamma_{\bf k}=(\cos k_x +\cos k_y)/2$,
which is
zero along the lines $(\pi,0)$-$(0,\pi)$. This gives a diamond-like
shape of $n_{\bf k}$. 
In the ADW configuration
$\langle\tilde{c}^\dag_i\tilde{c}_{i+d}\rangle$ for ${\bf d}=(2,0)$
and  $(0,2)$ are strongly suppressed  due to the
$\pi$-shift of the antiferromagnetic order parameter across the wall
and across the holon. As a result, there are significant and almost
equal 
$\cos k_{x,y}$ and  $\cos k_x \cos k_y$  harmonics
of opposite sign in $n_{\bf k}$ (see Table I), but the harmonics
$\cos 2 k_{x,y}$ are much suppressed. This is different from the
spin-polaron case where they are all of the same order.
Our results for $n_{\bf k}$ in Fig. \ref{fig_25} 
are obtained for the single-hole problem in a
stripe configuration, but 
since $n_{\bf k}$ is ``saturated'' on short distances the same
characteristic features should remain the same
for higher doping. In fact, our
``Maltese-cross''-like shape of $n_{\bf k}$ is remarkably close to that
observed in angle-resolved photoemission spectroscopy, 
Ref.~[\onlinecite{ARPES}].

Altogether, we have a very close agreement of the theory and DMRG data
on the energy, density, and electron distribution function for the
single excitation in the stripe configuration. This proves that our
description of such an excitation is correct.

\subsection{Many-hole problem}

The second problem we address in this work  is the many-hole system. 
As we described in Sec. \ref{Numerics} the finite cluster can be doped
with different amount of holes and one can use staggered-field BC's at
the open boundaries to enforce the state with and without an
ADW. At some doping concentration the stripe state with an ADW becomes
a ground-state and no fields are necessary to stabilize
it. We also note
that for two and more holes the M\"obius BC's are not beneficial anymore
because the holes play the role of boundaries for each other and the
meandering of the domain wall can come as a result of some
``collective'' motion. Our Fig. \ref{fig_26} shows  an example of
such a process for the case of two holes. Such effects were discussed
earlier in Refs.~[\onlinecite{Oleg_T,Zaan1}]. However, we will show
that the role of such processes in the stripe energy 
is negligible. This can be anticipated
since the holes are spread significantly within the individual
excitation and such collective
processes should be statistically rare. 
Moreover, the ``bending'' of the
stripe affects the free 
longitudinal motion of the holes and thus is unfavorable. 
This is in accord with the earlier work, Ref.~[\onlinecite{Wilczek}],
where such 
a rigidity was referred to as a ``garden hose'' effect.
Note that in
the case of a strictly 1D stripe close to complete filling
($n_{||}\simeq 1$) the
longitudinal kinetic energy is suppressed and the stripe 
meandering can become more important \cite{Zaan1}.
\begin{figure}[t]
\includegraphics[angle=270,width=3in,clip=true]{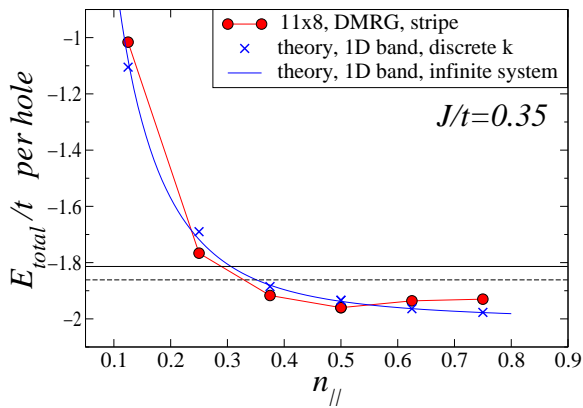}
\caption{Total energy of the system with an ADW per hole
versus $n_{\parallel}$. 
Circles are the DMRG results from the $11\times 8$ cluster.
Solid curve and crosses are the theoretical results as described in
the text.
Horizontal solid and dashed
lines are the energies of free spin polarons and bound states of spin
polarons in the homogeneous AF, respectively. $J/t=0.35$.}
\label{fig_27}
\end{figure}

\subsubsection{Total energy}

In Fig. \ref{fig_27} we show the DMRG and theory results for the total
energy of the system per hole versus linear hole concentration
$n_\parallel=N_h/L_y$ (energy of the empty system being subtracted).
DMRG data are obtained in the $11\times 8$ cluster with
$N_h=1,2,3,4,5$ and $6$ holes populating an ADW. At $N_h>3$ there
was no need to stabilize the stripe by the fields at the boundaries 
since it was the
ground state of the system. The theoretical curve is equivalent to 
the results shown in Fig. \ref{fig_9a}, which are obtained
from the rigid-band
filling of the effective 1D band using Eq. (\ref{E}). Note again that
the reference energy is different from Sec. \ref{Analytics} by $2J$. 
Crosses are also the result for the same rigid-band filling but for
the system with the discrete ${\bf k}$-space. We mimic the
periodic finite-size system by imposing that only $L_y=8$ ${\bf
k}$-points are available in the effective 1D band. Straight solid and
dashed lines are the energies  per hole 
of the systems of independent spin polarons and bound states of spin
polarons  in the homogeneous AF, respectively. 

Obviously, the rigid band filling
neglects the effects of interaction between the carriers except
the Fermi repulsion. Such interactions are quite complicated and 
would include attractive as well as repulsive terms as well as some
collective effects resulting in stripe meandering.
However, given the good 
agreement of analytical and numerical results one can conclude
that such effects are secondary for the stripe formation. Therefore,
the kinetic energy of the individual 
holes, both along the stripe (holonic
motion) and perpendicular to it (spin-polaron part), is the main 
reason that brings the stripe to the ground state.
Note that since the holons are spinless, the Fermi exclusion is much more
effective at inducing proper correlations between holons, 
acting as a hard core repulsion, than in the usual spin-1/2 case,
where up and down spin particles can be on the same site. Thus, the
effectiveness
of this simple band filling approach is not so surprising.

One can question the physical picture of a ``straight'',
weakly meandering  stripe described by an
effective 1D band from the point of view of applicability of
the ``free-holon'' approximation, Ref.~[\onlinecite{Oleg_T}]. The
controversy is that at the physically relevant concentration
$n_\parallel\simeq 0.5$ the free-holon approximation  
($n_\parallel\ll 1$) and the free-electron approximation 
($1-n_\parallel\ll 1$) are not applicable 
and should be equally bad. A very good
agreement of our theory with the numerical data
up to $n_\parallel \sim 0.5$ and beyond  can be seen as
quite puzzling since the theory  is based on 
$n_\parallel\ll 1$ approach. 
However, such a controversy comes from the
mean-field picture of the strictly 1D stripe.
One should rather consider a stripe 
to be a combination of strongly dressed, well spread
holes forming a collective bound state with the ADW. 
In fact, the actual amount of holons within the 
effective 1D band is given by $n_{holons}=\sum_{k_y<k_F} Z_{k_y}$ 
($ Z_{k_y}$ is a residue of the Green's function), which,
for the physical range of parameters, does not exceed
$n_{holons}\simeq 0.2 \ll 1$ even for a completely filled stripe, 
$n_\parallel=1$. 
It demonstrates that the $n_\parallel\ll 1$ approach should
work well in all ranges of doping.
\subsubsection{Chemical potential}

Figure \ref{fig_27} provided a comparison of 
the total energy of the system within the different topological
sectors, which defines the ground state. However,  
a more discrete energetic analysis is necessary 
to study the delicate balance of the stripe formation.
Here we introduce the chemical potential as the difference
between the energy of the system with $N_h$ and $N_h-1$ number of holes:
\begin{eqnarray}
\label{Eb}
\mu(N_h)=E^{N_h}_{tot}-E^{N_h-1}_{tot}\ .
\end{eqnarray}
In the situation when the kinetic energy is frustrated,
it gives a measure of how effectively the energy of the system  is
lowered by an extra hole for the states with different topology.

\begin{figure}[t]
\includegraphics[angle=270,width=3in,clip=true]{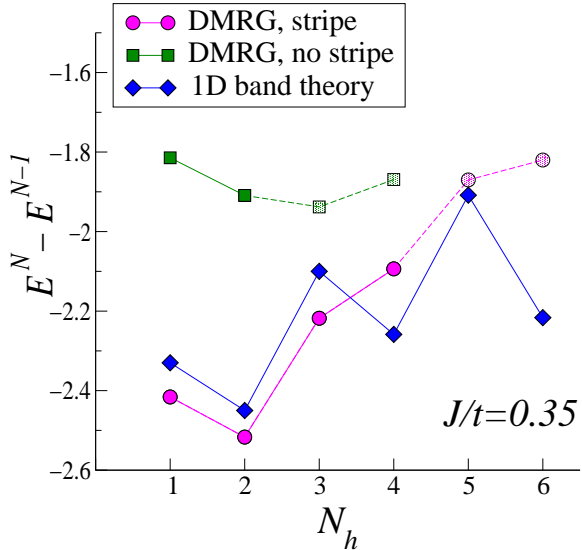}
\caption{Chemical potential vs $N_h$. Discrete-$k$ rigid-band
results, $L_y=8$, (diamonds), DMRG data in $11\times 8$ cluster for
the system with ADW (circles) and  
without ADW (squares), $J/t=0.35$. Lines are guides to the eye.}
\label{fig_28}
\end{figure}

Figure \ref{fig_28} shows 
$\mu$ as a function of $N_h$ for the DMRG 
in the $11\times 8$ cluster with cylindrical BC's for the
states with and without the stripe, together with theoretical
results calculated from the discrete-$k$ rigid-band filling of the 
effective 1D stripe band.
In obtaining theoretical data points we needed to account for the 
frustrating character of the cylindrical BC's 
(discussed above for  one hole) for the cases when the
number of holes is odd:
\begin{eqnarray}
\label{mu_th}
E_{tot}^{N_h}=\sum_{n=1}^{N_h}E_{kin}(k_n)+(3J/2) N_h
+(J/2)\delta_{N_h,odd} \ ,
\end{eqnarray}
where $k_n$ is one of the available $L_y=8$\ ${\bf k}$-points, counted
from the bottom of the band, $E_{kin}(k_n)$ is the energy of the 1D
excitation, Fig. \ref{fig_9}, \ $3J/2$ is the energy of the static hole
at the ADW, $J/2=E_{spinon}$ is the frustration energy caused by BC's. 
As a result, the theoretical expression for the chemical potential is 
given by:
\begin{eqnarray}
\label{mu_th1}
\mu^{th}(N_h)=E_{kin}(k_{N_h})+3J/2+(J/2) (-1)^{N_h} \ ,
\end{eqnarray}
where $E_{kin}(k_{N_h})$ is the lowest energy available for the $N_h$ 
hole in the 1D band. The last term provides a zigzag behavior of $\mu(N_h)$ 
shown in Fig. \ref{fig_28}.

For $N_h=1$ the DMRG and the theory points are equivalent to those
in Fig. \ref{fig_21} for the $L_y=8$ cluster with periodic BC's and 
for the $E_{th}^{GS}=E_{holon}+E_{spinon}$, respectively. Since the DMRG
data in Fig. \ref{fig_21} seem to scale to $E_{th}^{GS}$, the
difference $\mu^{DMRG}(1)-\mu^{theory}(1)$ is, most probably, a
finite-size effect.  
One can see a very good
overall agreement of the trends in both the numerical and analytical
result. Since the theoretical results are calculated from the 
picture which corresponds to the subsequent filling
of the 1D band, the chemical potential should grow as the
higher $k$-states are filled. 

\begin{figure}[t]
\includegraphics[width=3in,clip=true]{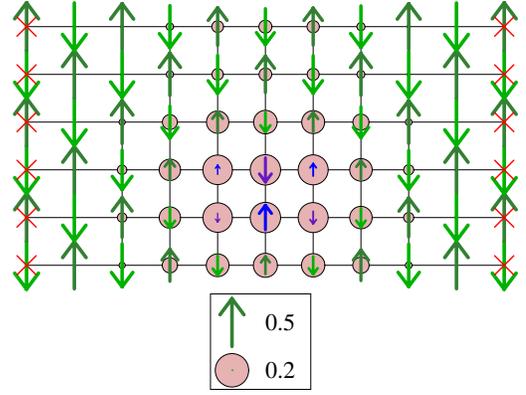}
\caption{$11\times 6$ cluster with four holes. 
``Homogeneous'' antiferromagnetic state is enforced by the boundary
conditions. Instead of being homogeneous, the circular ADW is created.}
\label{fig_29}
\end{figure}

We mark differently the data for 
$N_h=5$ and $N_h=6$ since in the $L_y=8$ stripe they 
correspond to the high concentrations where the finite-size effects
become more pronounced. 
Note again that the ADW configuration 
(stripe) is enforced by the boundary conditions at $N_h=1,2$
while for $N_h=4,5,6$ it is the ground state. Conversely, the no-stripe 
state (homogeneous AF) is the ground state for $N_h=1,2$ ($N_h=3$
looks like a metastable state) and is to be enforced for $N_h>2$. We
mark the no-stripe data for $N_h=3,4$ as shaded because the $N_h=4$ state
is not formed by the gas of spin polarons or polaron pairs but rather is 
a stripe-like circle with an ADW in the center of the cluster
(Fig. \ref{fig_29}).  
It is not a bound state of polaron pairs, but rather a many-particle
bound state with the condensate of magnons,  the ``droplet'' of the
$\pi$-shifted AF inside the circle corresponds to such a condensate. 
This
shows the strong tendency to the stripe formation such that 
even a small, 
finite number of holes will prefer  to  form a closed loop of the ADW 
``nuclei'', which can then develop into the straight stripes as the
doping grows.
We would like to discuss here the $N_h=1$ and $N_h=2$ cases in
Fig. \ref{fig_28} in more detail. 
While the no-stripe state is the ground state for these hole
concentrations the effectiveness of the energy lowering is much
higher in the stripe state. For $N_h=1$ it is yet another form of the
discussion given in Sec. \ref{Analytics} that at the bottom of the 1D
holon-spin-polaron band the energy is significantly lower than in the
bulk. In other words, from Fig. \ref{fig_28} one can see that the
individual charge carries benefit energetically from being at the
stripe.
Since for $N_h=1$ in either the stripe or the no-stripe system 
it is the kinetic energy of the individual charge 
excitation which is  optimized, one can conclude that the formation of 
the stripe is ``kinetic-energy driven''. 

The idea that stripe formation can be viewed as a condensation of 
a set of hole pairs has been discussed in Ref.~[\onlinecite{SD_pairs}].
Here we see that if one preconfigures the ADW, even a single hole
is bound to it with substantial binding energy, and pairing is
not involved. The quantitative description of stripe formation
that we have developed, which does not involve pairing, suggests
that pairing is a lower energy phenomena which can be considered
after the stripe is formed. However, the condensation of pairs
idea may also be a valid point of view. An energetic test for
this point of view would be that the energy per hole of a stripe
should be only slightly less than that of separate pairs. More
specifically, the energy advantage of the stripe should be less
than the pair binding energy.  For the $11\times 8$, $J/t=0.35$ system, 
we find that the energy of the half-filled stripe, per pair of holes,
is $2 E^{stripe}_{1/2}/N_h= -3.91$. This compares with 
$E_2=-3.72$ for a single pair. The difference between these two
energies is approximately twice the binding energy of a pair of holes,
$\Delta\simeq -0.09$,
suggesting that the condensation of hole pairs is not nearly
as good a description as the current formalism, at least
in the $t$-$J_z$ model.

A separate issue is whether the stripes help to promote pairing,
are irrelevant to it, or hinder it.
In the no-stripe state the  
chemical potential is lower for $N_h=2$ than for $N_h=1$. This energy
is the true bound-state energy of the pair of spin polarons,
$\Delta_{sp}=\mu(2)-\mu(1)<0$, much studied in the past,
Ref.~[\onlinecite{CP}]. In the stripe state the excitations do not
form a true bound state, although $\mu(2)-\mu(1)<0$. 
The theoretical result for this difference,
$\mu(2)-\mu(1)=E_{kin}(k_2)-E_{kin}(k_1)-J$, contains the negative
energy $-J$ provided by the removal of the spinon, which is 
induced by the PBCs in the one-hole system and has nothing to do with
the pairing. Since the DMRG data look very
much the same and also $\mu^{DMRG}(2)-\mu^{theory}(2)$ is almost the same
as for the one-hole case, one may conclude that there is no binding 
involved here at all. However, it may also be that the binding energy 
is compensating the stronger finite-size effects for the $N_h=2$ case
leading to the same $\mu^{DMRG}(2)-\mu^{theory}(2)$.
In any case, we see no significant enhancement
of binding and thus the stripes seem to be largely irrelevant to
pair-binding.

The collective stripe fluctuations (as opposed to the fluctuations of
individual holes) have been discussed as an effective 
alternative way to lower the energy and stabilize the stripe phase 
\cite{Zaan1}. We check the effectiveness of such processes by studying
two holes in the $11\times 8$ system with periodic BC's and M\"obius
BCs. The M\"obius BC's suppress 
the stripe meandering considerably, which is seen in the hole density
profiles, while the difference between the energies of the states is
slightly above the numerical accuracy of the DMRG.

The overall conclusion of this discussion is the following. The
kinetic energy of the {\it individual} charge excitations at the
domain wall, which includes significant components of both
``longitudinal'' and ``transverse'' motion, is the reason for the
stripe formation. The pairing energy does not seem to be significantly
modified and, in general, is associated with the smaller energy
scale. Meandering of the stripe, while being important for some 
other properties, has very little effect on the energy of the stripe.
Therefore, one can write this as a hierarchy of the energy scales:
\begin{eqnarray}
\label{E_hi}
\Delta E_{kin}\gg E_{pairing}\gg E_{meandering}\ ,
\end{eqnarray}
where $\Delta E_{kin}$ is the {\it difference} in the ``release'' of
the kinetic energies between the stripe and homogeneous state,
$\Delta E_{kin}\sim (J^2t)^{1/3}$, as discussed in Sec. \ref{Analytics}. 
$E_{pairing}\sim J$ but, in fact, is only a fraction of $J$,
Ref.~[\onlinecite{CP}]. $E_{meandering}$ should carry a statistically
small factor describing the probability of the collective motion of
two or more holes together. As we mentioned, $E_{meandering}$ is
hardly detectable numerically.

The emerging picture of the stripe as a collective bound state of 
strongly dressed 1D band excitations with the ADW 
also implies that there are
``deep'' states, which reduce the energy of the stripe, and
``shallow'' states, which are spread around the stripe and are 
only weakly coupled to it. 

\subsubsection{Density}

\begin{figure}[t]
\includegraphics[angle=270,width=3in,clip=true]{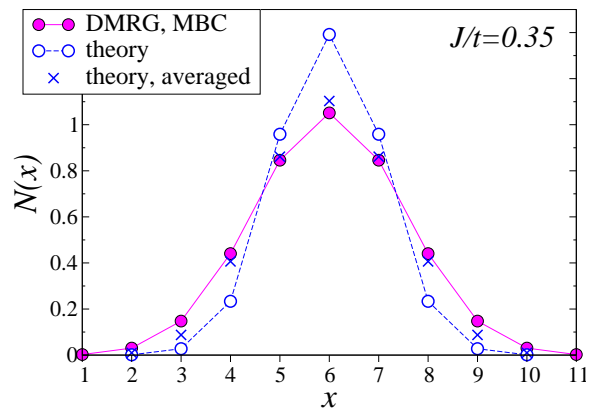}
\caption{Hole density distribution across the stripe at $J/t=0.35$. 
DMRG data for four holes in $11\times 8$ cluster, 
cylindrical BC's (filled circles). Theoretical results for (i) a 
half-filled stripe centered at $x_0=6$ (empty circles); (ii) linear
combination of stripes at $x_0=5$, $x_0=6$, and $x_0=7$
(crosses) are shown.
Lines are guides to the eye.}
\label{fig_30}
\end{figure}

To conclude this section we show the hole density profile for the
half-filled stripe in the $11\times 8$ cluster (four holes) compared to the
theoretical results for $n_\parallel =1/2$ from
Eqs. (\ref{N_x})-(\ref{wf}), $J/t=0.35$, Fig. \ref{fig_30}. In DMRG
data cylindrical BC's are used and the stripe is the ground state for
this system. Theoretical results for the 1D-band
stripe centered in the middle of the system at $x_0=6$ are shown by 
the empty circles. As we
discussed before, the meandering effect causes the migration of the
stripe as a whole in the transverse direction. Since we know that the
meandering effect is weak and that it effectively leads to the
coupling of stripes centered at the different $x_0$ we, therefore,
model this effect by assuming that the ground state is given by the
linear superposition of 1D bands centered at $x=5$, $6$, and $7$ with
equal weight. Further distribution is assumed to be unfavorable
because of the open BC's. Then the density profile is given by the
average:
$\bar{N}(x)=[N_{x_0=5}(x)+N_{x_0=6}(x)+N_{x_0=7}(x)]/3$.
The results of such an averaging are also shown in Fig. \ref{fig_30}
(crosses). Even better agreement can be reached assuming wider
meandering and weight distribution, but such a task is beyond the scope
of this work.

\section{Conclusions}
\label{conclusions}

Summarizing, we have presented a comprehensive
comparison of the DMRG numerical data 
for clusters up to $11\times 8$ 
with the analytical studies based on the 
self-consistent Green's function method
for a single stripe of holes in an AF described by the
$t$-$J_z$ model. We consider the close agreement of the results as a strong 
support for the validity of our analytical method and of the physical
picture which follows from it.
We have provided a description
of the charge carriers building the stripe as a system of 1D
elementary excitations, unifying the features of holons 
and antiferromagnetic
spin polarons. Then the stripe should be seen as an 
effective 1D band partly filled with these elementary excitations. 
This picture is in a very good accord with the numerical data.

As it follows from our study, the stripe can be roughly described by
the deep ``backbone'' states, which minimize the energy of the
anti-phase configuration in the AF and the shallow, almost free
spin-polaron-like excitations around the ADW. Since the spin polarons
are known to have a considerable pairing between themselves, such a
framework does not require the superconducting pairing to come from
some 1D instability, but rather suggests that the pairing is largely
unrelated to the 1D stripe pattern. Such a scenario is also discussed
in recent work, Ref.~[\onlinecite{Antonio_alone}]. 
Another hypothetical advantage of our picture is a more effective
screening of the long-range component of the Coulomb repulsion, which
represents the problem for the system of strictly 1D
charges.\cite{Kabanov} 
  
Altogether, the comprehensive comparison of the results of the theory and
DMRG numerical approach has shown a very close quantitative agreement, thus
providing a strong support to our way of understanding the charge
excitations at the anti-phase stripe in an AF.

\begin{acknowledgments}

We would like to acknowledge invaluable discussions with A.~Abanov,
E.~Dagotto, L.~Pryadko, O.~Tchernyshyov, J.~Tranquada, and S.~Trugman.
We are indebted to A.~Bishop for numerous stimulating conversations.
This research was supported in part by 
Oak Ridge National Laboratory, 
managed by UT-Battelle, LLC, for the U.S. Department of Energy under
contract No. DE-AC05-00OR22725, and by a
CULAR research grant under the auspices of the US Department of Energy.
SRW acknowledges the support of the NSF through grant DMR 98-70930.

\end{acknowledgments}



\begin{thebibliography}{}
\bibitem{remark} The problem of the stripe stability to the untwist of the
order parameter has been studied as well, Ref.~[\onlinecite{CCNB}].
\end{thebibliography}
\end{document}